\documentstyle{mn}
\begin{document}
\input{psfig}
\title[CLASS lens candidates]{The Cosmic Lens All-Sky Survey:II.
Gravitational lens candidate selection and follow-up.}
\author[I.W.A. Browne et al.]{I W.A. Browne$^1$, 
P.N. Wilkinson$^1$, N.J.F. Jackson$^1$, S.T. Myers$^{2,4,5}$,
\newauthor C.D. Fassnacht$^{4,2,8}$, L.V.E. Koopmans$^{6,1,4}$,
D.R. Marlow$^{1,5}$, M. Norbury$^1$, D. Rusin$^{5,9}$, \newauthor
C.M. Sykes$^1$, A.D. Biggs$^1$, R.D. Blandford$^{4}$, A. G. de
Bruyn$^{3}$, K-H. Chae$^1$, \newauthor P. Helbig$^{1,6}$,
L.J. King$^{7}$, J. P. McKean$^1$, T.J. Pearson$^{4}$,
P.M. Phillips$^1$,\newauthor A.C.S. Readhead$^{4}$,
E. Xanthopoulos$^1$, T. York$^1$\\ $^1$University of Manchester,
Jodrell Bank Observatory, Nr. Macclesfield, Cheshire SK11 9DL\\
$^{2}$National Radio Astronomy Observatory, P.O. Box 0, Socorro, NM
87801, USA\\ $^{3}$NFRA, Postbus 2, 7990 A A Dwingeloo, The
Netherlands\\ $^{4}$California Institute of Technology, Pasadena, CA
91125, USA \\ $^{5}$Dept. of Physics and Astronomy University of
Pennsylvania 209 S. 33rd Street Philadelphia, PA 19104, USA\\
$^{6}$Kapteyn Astronomical Institute, Postbus 800, 9700 AA Groningen,
Netherlands \\ $^{7}$ University of Bonn, Auf dem H\"{u}gel 71,
D-53121 Bonn, Germany\\ $^{8}$ STScI, 3700 San Martin Dr., Baltimore,
MD 21218, USA \\ $^{9}$Harvard-Smithsonian Center for Astrophysics, 60 Garden Street, Cambridge, MA02138.}

\date{October 18, 2002}

\maketitle

\begin{abstract}
We report the final results of the search for gravitationally lensed
flat-spectrum radio sources found in the combination of CLASS (Cosmic
Lens All-Sky Survey) and JVAS (Jodrell-Bank VLA Astrometric
Survey). VLA observations of 16,503 sources have been made, resulting
in the largest sample of arcsec-scale lens systems
available. Contained within the 16,503 sources is a complete sample of
11,685 sources having two-point spectral indices between 1.4 and 5~GHz
flatter than -0.5 and 5~GHz flux densities $\geq$30~mJy.  A subset of
8,958 sources form a well-defined statistical sample suitable for
analysis of the lens statistics.  We describe the systematic process
by which 149 candidate lensed sources were picked from the statistical
sample on the basis of possessing multiple compact components in the
0.2~arcsec-resolution VLA maps. Candidates were followed up with
0.05~arcsec resolution MERLIN and 0.003~arcsec VLBA observations at
5~GHz and rejected as lens systems if they failed well-defined surface
brightness and/or morphological tests.  To illustrate the candidate
elimination process, we show examples of sources representative of
particular morphologies that have been ruled out by the follow--up
observations.  One hundred and ninety four additional candidates, not
in the well-defined sample, were also followed up. Maps for all the
candidates can be found on the World Wide Web at
http://www.jb.man.ac.uk/research/gravlens/index.html. We summarize the
properties of each of the 22 gravitational lens systems in
JVAS/CLASS. Twelve are double-image systems, nine are four-image
systems and one is a six-image system. Thirteen constitute a
statistically well-defined sample giving a point-source lensing rate
of 1:690$\pm$190. The interpretation of the results in terms of the
properties of the lensing galaxy population and cosmological
parameters will be published elsewhere.

\end{abstract}

\section{Introduction}

Gravitational lens systems play a vital role in the investigation of
the distribution of matter in individual galaxies and clusters
(Kochanek, 1993; Kochanek, 1995; Rusin \& Ma, 2001; Keeton \& Madau,
2001; Koopmans \& Treu, 2002) and in the determination of the Hubble constant, $H_{0}$
(Refsdal 1964), while statistical analyses of gravitational lens
surveys can place  constraints on the cosmological constant,
$\Lambda$, and the matter density, $\Omega$ [eg Turner 1990; Fukugita
et al., 1992; Carroll, Press \& Turner 1992; Kochanek 1996; Falco,
Kochanek \& Mun{\~o}z 1998; Helbig et al., 1999; Macias-Perez, et al.,
2000]. When using lens statistics for cosmology a major concern is the
statistical completeness of the samples. For example, in optical
searches, lens systems can be missed if the lensing galaxy is bright
compared to the lensed images or if extinction within the lens hides
lensed images.  Similarly, in radio searches, the existence of
extended radio structure in the lensed source on the scale of the
expected multiple imaging can make lensing events difficult to
identify reliably. Moreover, extended radio structure makes the
lensing probabilities difficult to assess since they depend both on
the lens cross-section and on the intrinsic extent of the radio source
being imaged. Creating a reliable and complete sample of
multiple-image lens systems is not easy.

There are several ongoing or completed surveys for arcsecond--scale
(i.e. individual galaxy mass) gravitational lenses that have led to
the discovery of multiple--image systems\footnote{see the web site of
the CfA Arizona Space Telescope Lens Survey
(http://cfa-www.harvard.edu/castles/) for a summary of the known
multiply-imaged systems and their properties}.  The largest systematic
{\it optical} search was the HST Snapshot survey, which detected 5
lens systems in a sample of 502 highly luminous, quasars (e.g. Bahcall
et al., 1992, Maoz et al., 1993). The first large scale {\it radio}
survey was the MG-VLA survey (see Burke et al. 1993, Hewitt et al.,
1992 for details), which also discovered five gravitational lens
systems. Since there was no radio-spectral index selection imposed on
targets in the MG-VLA sample, some lensing events may have been missed
because of the difficulty of recognizing multiply-imaged single
sources in a sample in which many of the steep-spectrum sources have
intrinsically complex structures (Kochanek \& Lawrence 1990). However,
such surveys do find examples where radio lobes are lensed into rings
and this has been exploited systematically by Leh\'{a}r et al. (2001).

In contrast to the MG-VLA survey, the $\sim$2500 source Jodrell-Bank
VLA Astrometric Survey (JVAS; Patnaik et al., 1992a, Browne et al.,
1998, Wilkinson et al., 1998; King et al., 1999) was limited to
targets with flat radio spectra ($\alpha\geq-0.5$ where
$S\propto\nu^{\alpha}$).  This approach has several advantages:

\begin{itemize}

\item Flat-spectrum radio sources have intrinsic structures dominated by a
single milliarcsec radio ``core''. Thus a source found to have
multiple compact components, when observed with $\sim$200~mas
resolution, is automatically a strong lens candidate

\item Higher-resolution radio  mapping of lens
candidates with MERLIN and VLBI is a reliable way to discriminate between
intrinsic structure and multiply-imaged cores.

\item The probability of multiple imaging of radio cores of angular
extent $\ll$ the lens Einstein radius depends only on the lens
cross-section and not on the angular extent of the lensed object.

\item Many flat-spectrum radio sources are quasars and therefore at high 
redshift, thus maximizing the lensing probability.

\item Most flat-spectrum radio sources are time variable and, if
multiply-imaged, are suitable for time delay measurements and hence
Hubble constant determination.

\item Any bias arising from extinction in the lensing galaxy, a major
concern in assessing the completeness of optical lens searches, is
eliminated.

\item Any radio emission from the lensing galaxy is likely to be faint
and thus confusion between lens and source emission is unlikely
(in contrast to the situation at optical wavelengths).

\item MERLIN and VLBI maps often reveal sub-structure within the
lensed images on a scale $\le$ the Einstein radius which provides extra
constraints on the lens mass-model.

\end{itemize}

The JVAS survey led to the discovery of six lens systems, five new
ones and one rediscovery, (King et al., 1999). The Cosmic Lens All Sky
Survey (CLASS) described here used the JVAS approach and has extended
it to much larger numbers of targets and to weaker radio sources.

In order to work on a much larger sample of objects, a collaboration
was initially formed between the University of Manchester group at the
Jodrell Bank Observatory (JBO) and groups from the California
Institute of Technology and in the Netherlands (Dwingeloo and Leiden
University). The CLASS team now comprises  groups at JBO, California
Institute of Technology, NRAO, STScI, CfA and Dwingeloo. CLASS has
concentrated on the Northern hemisphere; a survey with similar goals is being
pursued by others in the South (Winn et al., 2000)

\begin{figure*}
\centering
\setlength{\unitlength}{1in}
\begin{picture}(7.5,6.5)
\put(1.6,0.3){\includegraphics{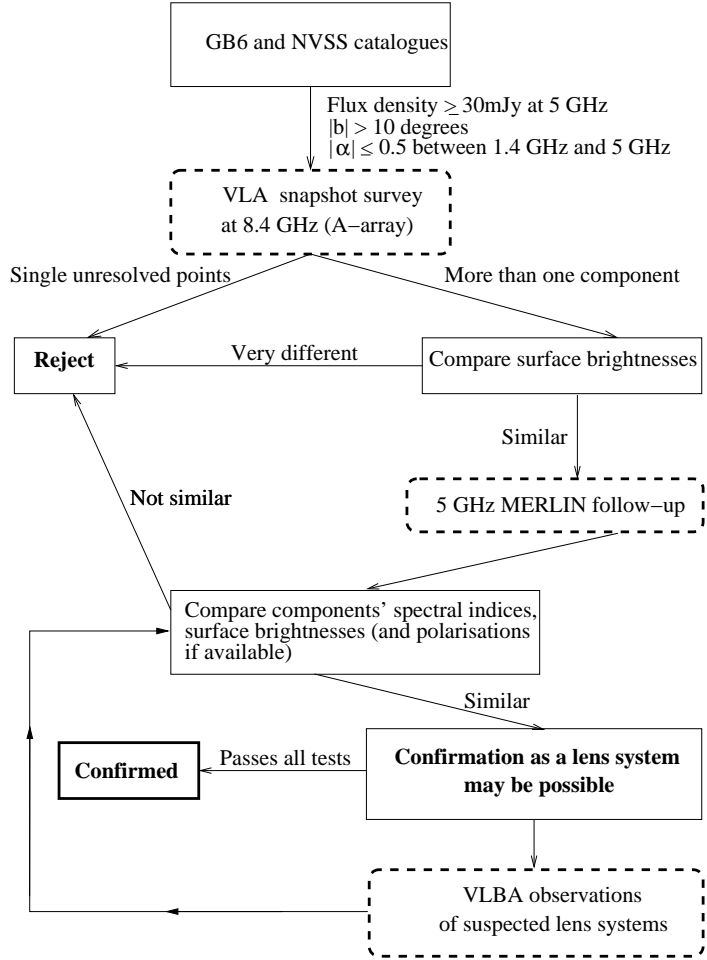}}
\end{picture}
\caption{Flow diagram illustrating the fundamentals of the CLASS
search strategy. In some individual cases extra VLA, MERLIN and VLBI
observations are made where necessary. Also optical observations were
attempted for those candidates that looked very promising. No
candidates were rulled out on the basis of optical observations
alone. This methodology is very similar to that adopted for the JVAS
search (King et al., 1999).\label{flow}}
\end{figure*}

The selection of targets, the VLA observations and their analysis are
described in full in the companion paper, Myers et al. (2002),
hereinafter Paper 1. In the present paper we concentrate on the
selection of the lens candidates and the follow up procedures; these
procedures are very similar to those used for JVAS and illustrated in
Figure~1 of King et al. (1999). That figure is reproduced here in slightly
modified form (Figure~\ref{flow}). The candidate selection, and the
subsequent rejection of most of them, is discussed in some detail to
validate our claim that JVAS/CLASS have produced reliable and complete
lens statistics which can then be used to constrain the parameters of
cosmological models (Chae et al., 2002; hereinafter Paper 3). We list the candidates which have been followed
up and give the web address  (http://www.jb.man.ac.uk/research/gravlens/index.html)
where the MERLIN and VLBA maps can be found and examined.

\section{Primary Sample Selection.}

Paper 1 gives full details of source selection and data analysis; here
we give only an abbreviated account for completeness.  The final CLASS
complete sample was selected from the 5~GHz GB6 catalogue (Gregory et
al., 1996) and the NVSS 1.4~GHz catalogue (Condon et al., 1998).
Starting with GB6 we picked sources stronger than 30~mJy and looked
for their counterparts in NVSS, rejecting all those with two point
spectral indices steeper than -0.5.  The area covered by GB6 is from
declination 0$^{\circ}$ to $75^{\circ}$. To avoid difficulties with
Galactic extinction in optical follow up observations or confusion
with extended Galactic sources like planetary nebulae, only sources
with $|\rm b|\geq 10^{\circ}$ were included.  A total of 11,685 GB6
sources meet these selection criteria of which all but 13 have been
observed successfully\footnote{Since the GB6 and NVSS surveys did not exist at the
outset of the CLASS project, the first target sources were selected
using slightly different criteria based on various combination of 87GB
(Gregory \& Condon, 1991), the WENSS 325~MHz survey (Rengelink et al.,
1997), the 365~MHz Texas survey (Douglas et al., 1996) and the NRAO
1.4~GHz survey White and Becker (1992). This means that during the
different stages of the CLASS project a significant number of sources
not in the final complete sample were observed resulting in a total of
16,503 sources being observed with the VLA (See Paper 1 for more
details).}. The 11,685 sources include $\sim$2000 JVAS sources and
henceforth the ``CLASS complete sample'' should be understood to
contain JVAS sources as a subset.  Each source selected was observed
with the VLA in its A-configuration at a frequency of 8.4~GHz resulting
in a map with a resolution of 200~mas, an rms noise level of
$\sim$0.2~mJy and a typical dynamic range of $\geq50:1$.  Even though
the CLASS data have been collected and reduced over a period of many
years, all have been been recently re-edited, re-calibrated and
re-mapped in a standard manner to ensure uniformity of data
quality. Further details of the VLA observations and their analysis
are given in Paper 1.

\section{Lens Candidate Selection.}

We looked for lens systems in which there are multiple images of a
compact, flat radio spectrum, core of a quasar or a radio galaxy.  The
crucial step in the whole survey process is the selection of
candidates for higher-resolution follow up observations. The method is
then to use the higher resolution maps to distinguish multiply-imaged
compact cores from more extended, non-lensed, components, the
assumption being that no radio sources have more than one compact VLBI
core within a few arcsec\footnote{The fact that we only have one
possible candidate ``dark lens'' (see Section 6.3) supports out
assumption}. The whole candidate selection and follow-up process is
illustrated in Figure~\ref{flow}. The CLASS philosophy has been to
follow up all objects meeting the well-defined, and cautious,
selection criteria listed below, even when our initial impression from
the VLA maps was that they were very unlikely to be multiply imaged
systems. It is because of this meticulous follow-up procedure that we
believe that our lens sample is complete within the selection
criteria.  We reiterate that for our statistically complete sample we
are searching only for multiply imaged compact radio cores\footnote{We
did not, of course, ignore obvious lens systems in which lensed
extended emission dominates; e.g 1938+666. Such objects should not,
however, be used in a statistical analysis of lensing rates since the
lensing probability depends both on the lens cross section and the
intrinsic size of the lensed object.}. For candidates to be included
in the systematic follow up they must meet the following criteria:

\begin{itemize}

\item The source must have multiple compact components each with a gaussian
diameter (FWHM) $\leq$170~mas when observed with the VLA at a
resolution of 200~mas (but see below).

\item The separation of the compact components  must be $\geq$300~mas; the search is
currently complete out to a radius 15~arcsec from the strongest component in the
map. Results of the search in the separation
range 6 to 15~arcsec have already been presented by Phillips et al (2001).

\item The component flux density ratio must be $\leq$10:1.

\item The sum of the 8.4~GHz flux densities of the components must be $\geq$ 20~mJy.

\end{itemize}

\begin{figure*}
\centering
\setlength{\unitlength}{1in}
\begin{picture}(7.5,6.5)
\put(1.6,0.3){\includegraphics{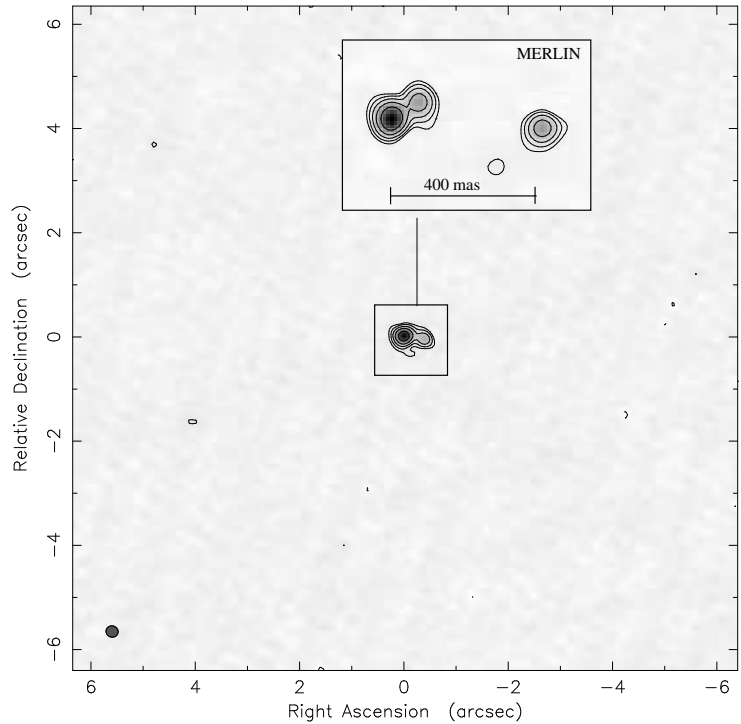}}
\end{picture}
\caption{The VLA 8.4~GHz finding map of the compact 4-image lens
system CLASS B1555+375 with the higher resolution MERLIN 5~GHz map
shown as an inset. Note how in the VLA map the 3  images merge
to resemble a double source with an extended core or a core with a
compact jet. The map shown in Figure~\ref{merlin2} shows all four images of the quad.\label{1555}}
\end{figure*}

The precise criteria adopted above were dictated by the instrumental
resolution and sensitivity of the VLA (see Paper 1) and the need to be
confident that the selection of candidates was both reliable and
complete within well-defined limits. Four-image lens systems, at least
those with separations $\geq$0.6~arcsec, are easier to identify than
two-image lens systems, since quads have more eye-catching
configurations that are highly unlikely to be intrinsic to the target
source. The majority of candidates have two components and these
candidates required the application of our rigorous follow-up protocol
to determine whether or not they were lensed images of a single
source. For characteristic separations $\leq$0.6~arcsec, however,
quads can mimic doubles when observed with 0.2~arcsec resolution. This
is illustrated by the CLASS B1555+375 system (Figure~\ref{1555};
Marlow et al., 1999a) where two merging images look like a single
strong and extended primary component, the third image is just
detected while the fourth image revealed by MERLIN (see
Figure~\ref{merlin2}) is too weak to show up on the original VLA
finding observations. For this reason we have taken special care and
followed up all sources which resemble B1555+375. In particular, we
have followed up 27 small separation doubles in which the strongest
component does not formally satisfy the angular size $\leq$170~mas
condition. None of these extra candidates turned out to be lens
systems.

Because of the adoption of the above criteria, the effective size of
the sample from which lens candidates are selected is significantly
less than the 11,685 sources in the CLASS complete
sample. In Table~\ref{numbers} we give a breakdown of  the source numbers. 
{\it The ``CLASS statistical sample''  which
should be used in statistical analyses, contains 8,958 objects.} The
present paper concentrates on the follow-up of 149 candidates
selected from the 8,958 sources in the CLASS statistical
sample. These 149 candidates are listed in Table~\ref{cand}. The
resulting sample of 13 lens systems will be referred to as the
``statistically well-defined lens sample''. Other sources not in the
complete sample, but which were observed with the VLA as part of
CLASS, have been treated in much the same manner for candidate
selection and a further 194 candidates were followed up. Nine new
lens systems have been discovered in this way (see
Table~\ref{additional}). In summary, we have followed up 343 (2.1\%)
of the 16,503 sources observed with the VLA.  

\begin{table*}
\caption{Breakdown of the numbers of sources in CLASS. The ``bandwidth
smeared'' sources refer to a number of, mostly JVAS, observations
where the VLA pointing position was more than one arcmin from the true
position of the source. Under these circumstances compact components
will appear artificially extended and can confuse the recognition of
lens candidates.
\label{numbers}}
\begin{tabular}{ll}
\hline  & Number of sources \\ \hline 
Total number of pointings & 16,503 \\
The CLASS complete sample & 11,685 \\
Failed observations & 13  \\
8.4~GHz flux density $\leq$20~mJy & 2,418 \\
Bandwidth smeared & 217 \\
Primary component $\geq$170~mas & 81 \\
The CLASS statistical sample & 8,958 \\ \hline
\end{tabular}
\end{table*}

Lens candidates were selected in two ways, first by visual inspection
of the VLA maps in combination with the analysis of the
automatically-generated results of the modelling of each source using
the Caltech DIFference MAPping package (DIFMAP; Shepherd, 1997); the
models were checked to make sure that the source met the selection
criteria listed above (see Paper 1 for more details of the model
fitting). This process was repeated several times, and by different
members of the CLASS collaboration, for the entire survey. Secondly a
list of candidates was also generated automatically from the DIFMAP
model-fit parameters using a figure-of-merit based on how well the
model parameters conformed to the selection criteria. This second
process served to cross-check the manual selection and to guard
against missing candidates due to book-keeping errors. All confirmed
lens systems had a figure-of-merit well above the limit that triggered
follow-up observations except for the very compact 4-image systems
B0128+437 and B1555+375. As we have mentioned above, special
precautions were taken not to miss similar systems. 

\begin{figure*}
\centering
\setlength{\unitlength}{1in}
\begin{picture}(11.5,6.5)
\put(1.6,0.3){\includegraphics{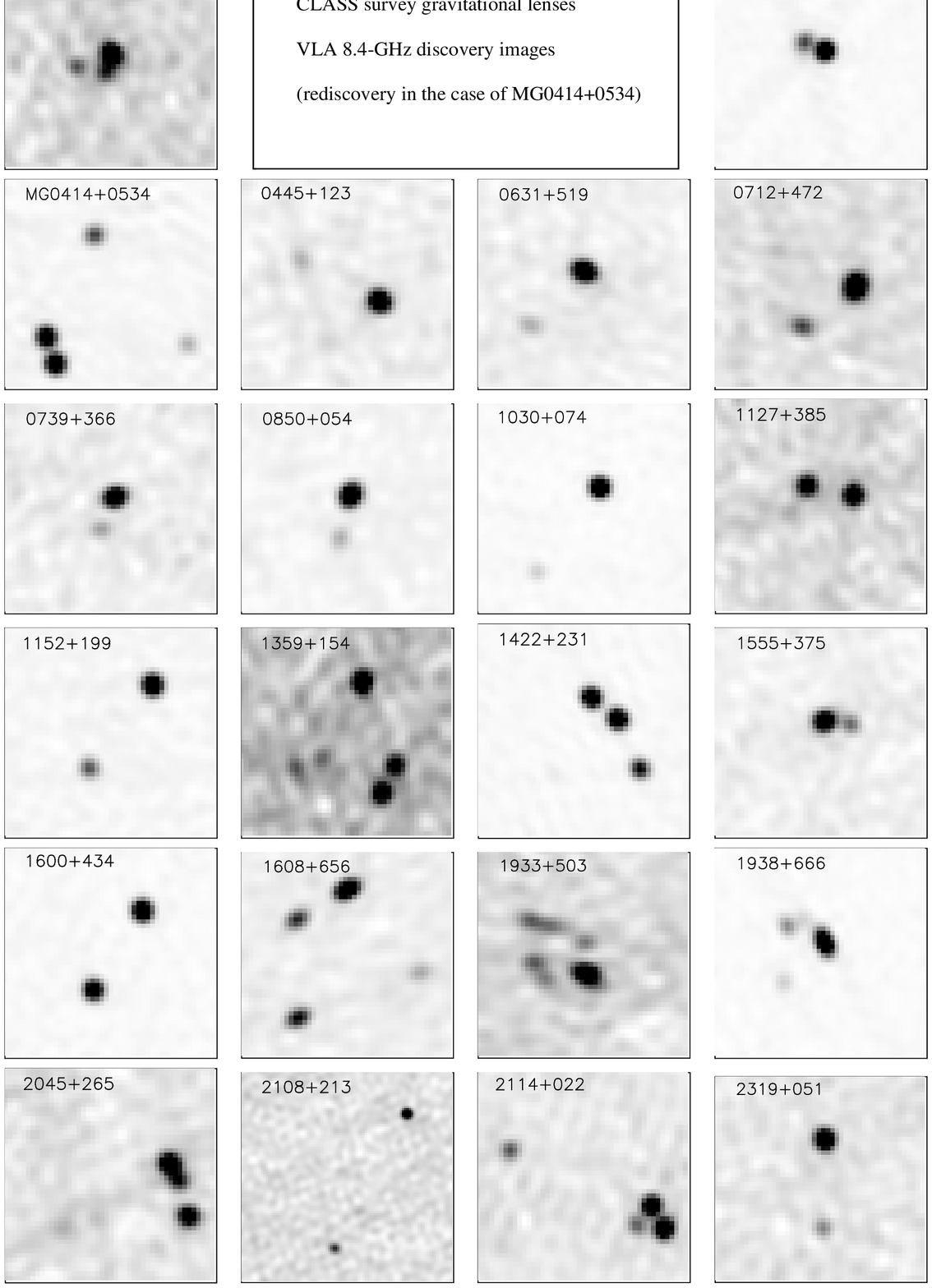}}
\end{picture}
\caption{VLA 8.4~GHz discovery  maps of the 22 successful lens candidates. \label{vla}}
\end{figure*}

The VLA discovery maps of the successful candidates are shown in
Figure~\ref{vla}. Some unsuccessful candidates are discussed in the
next section.

\section{Candidate Follow--up}

As illustrated in Figure~\ref{flow}, we followed up candidates with
stages of spatial filtering, systematically increasing the radio
resolution until all candidates could be classified. Since we expect
lensed images of the same object to have the same surface
brightness\footnote{Multipath scattering occurring in the
inter-stellar medium of the lens can change the measured surface
brightness as in for example CLASS B1933+503 (Marlow et al.,
1999b). However, scattering is a very strong function of observing
frequency and is unlikely to be a problem at frequencies of 5~GHz or
higher.} spectra and percentage polarization, lens candidates were
rejected if:

\begin{itemize}

\item The surface brightnesses of the putative lensed images were
different by more than a (conservative) factor of $\sim$4.

\item The structure revealed by the high resolution maps was
inconsistent with lensing; for example, the radio map showed two lobes
and a bridge of emission joining them, or a core plus a resolved knot
in a faint underlying jet.

\item The spectral indices  of the putative images and/or their
percentage polarizations at frequencies $\geq$8.4~GHz were very different (see below).

\end{itemize}

We were conscious that for multiply-imaged variable sources the
effects of time delays could make the spectral indices and
polarizations of the images of the same core appear different. In
addition depolarization and Faraday rotation in the ISM of the lensing
galaxy could affect the polarization properties of the images
differently. Thus we treated these quantities with caution when it
came to rejecting candidates. Only if one component had an obviously
steep spectrum, and therefore was very unlikely to be variable, did we
reject a candidate on the basis of radio spectral index
alone. Likewise we disregarded percentage polarizations at frequencies
$<$8.4~GHz where Faraday depolarization may be important. Even at
8.4~GHz and above we only rejected on the basis of percentage
polarizations if the component polarization differed by more than a
factor of four.

The first step in the follow-up path for all candidates was a snapshot
observation with MERLIN at 5~GHz. The typical time spent on each
target was one hour with observations made at four or five widely
spaced hour angles. These observations enabled a map to be produced
with a resolution of 50~mas and a typical rms noise level of
$\sim$0.2~mJy. In about 80\% of the cases the maps were enough to
reject unambiguously the VLA-derived candidate. In the majority of
these cases the extra resolution was sufficient to distinguish between
the components on the grounds of different surface brightness. This
was almost always in the sense that the secondary was resolved while
the primary remained compact. Since the secondaries are usually
several times weaker than the primaries this implies a large
difference in surface brightness. From the MERLIN 5~GHz results
components were adjudged to be resolved if they had FWHM $\geq$40~mas.
For some objects additional MERLIN observations at 1.7~GHz were made
and for others, mostly systems with components separated by more than
6~arcsec, multi-frequency (1.4, 5, 8.4, 15 and 22~GHz) VLA
observations were made (Phillips et al., 2001).

All candidates surviving the MERLIN filter were then observed with the
VLBA at 5~GHz giving maps with a resolution $\sim$3~mas and an rms
noise level of $\sim$0.15~mJy. The VLBA observations were again made
in snapshot mode and taken at four different hour angles.  By this
stage all but a handful of objects could be classified unambiguously;
the majority could either rejected on the basis of the criteria set
out above or  shown to satisfy those criteria. All but three
of the remaining ambiguous candidates were finally classified on the
basis of additional radio observations at different frequencies.
Those objects which passed all the radio tests were regarded as
``provisionally confirmed'' lens systems.  

The emphasis of the follow-up then shifted from confirmation to
consolidation.  We sought optical and/or infrared observations to try
to detect the lensing galaxy, the lensed images, and to measure their
redshifts. Imaging with the HST and spectroscopy with the Keck
Telescope proved a powerful combination. In all cases but one, that of
J0831+524 (see below), these observations of the provisionally
confirmed lens systems strongly supported the conclusion reached on
the basis of the radio data alone. Of the 23 candidates found to have two
or more milliarcsecond flat-spectrum radio components in the
separation range 0.3 to 6 arcsec, 22 are confirmed lens systems.

The results of the follow-up of the 149 candidates in the
statistically well-defined sample are summarised in
Table~\ref{cand}. For completeness it contains entries for those JVAS
sources belonging to the sample (marked with a $^*$) already reported
by King et al. (1999). For the same reason, some of the 6 to 15~arcsec
lens candidates (marked with a $^\dagger$) reported by Phillips et
al. (2001) are included. The reasons why candidates were rejected are indicated in columns 4 to 7.

\subsection{Discussion of individual candidates}

In this section we illustrate the candidate follow-up procedure with
some examples. We start with three simple cases in which rejection of
the candidate was straightforward after only one or two sets of
follow-up observations. We then discuss in some detail three of the
candidates which required extensive observations and still cannot
quite be ruled out with one hundred percent certainty.  The majority
of candidates were rejected because the surface brightnesses as
revealed by the high resolution maps of the putative lensed images
were different.
Sometimes, but not always, the conclusion to reject on the basis of
surface brightness was reinforced by the secondary structure having a
jet-like morphology.  Other structures which are sometimes seen are
Compact Symmetric Objects (CSOs) (Wilkinson et al., 1994; Fanti et
al., 1995, Readhead et al,. 1996) or Medium Symmetric Objects (MSOs)
(Augusto et al., 1998).

\subsubsection{Examples of candidates where  rejection is straightforward}

\begin{figure*}
\centering
\setlength{\unitlength}{1in}
\begin{picture}(7.5,6.5)
\put(1.6,0.3){\includegraphics{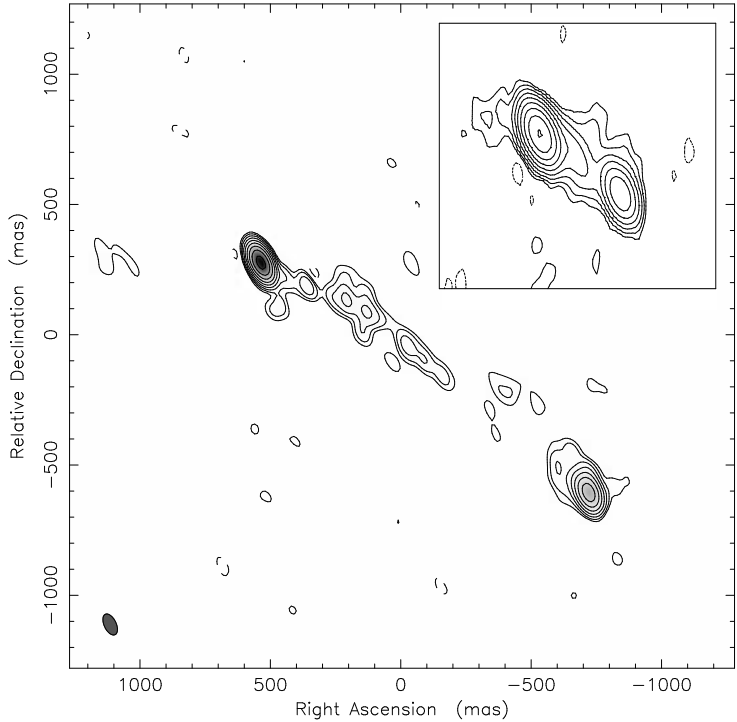}}
\end{picture}
\caption{MERLIN 5~GHz map of the rejected lens candidate J0448+098. The VLA 8.4~GHz map is inset for comparison \label{0448_merlin}}
\end{figure*}

{\bf J0448+098} Two well-separated and compact components of almost
equal flux-density are seen in the VLA finding map. A MERLIN 5~GHz map
(Figure~\ref{0448_merlin}) reveals low surface brightness emission
lying between the two main components suggesting that this is a MSO
or, just possibly a one-sided 3C273-like jet. Certainly the lensing
hypothesis can be rejected for this system because of the low
brightness emission between the more compact outer components which
could not be present if the outer components were lensed images of
each other.

\begin{figure*}
\centering
\setlength{\unitlength}{1in}
\begin{picture}(7.5,6.5)
\put(1.6,0.3){\includegraphics{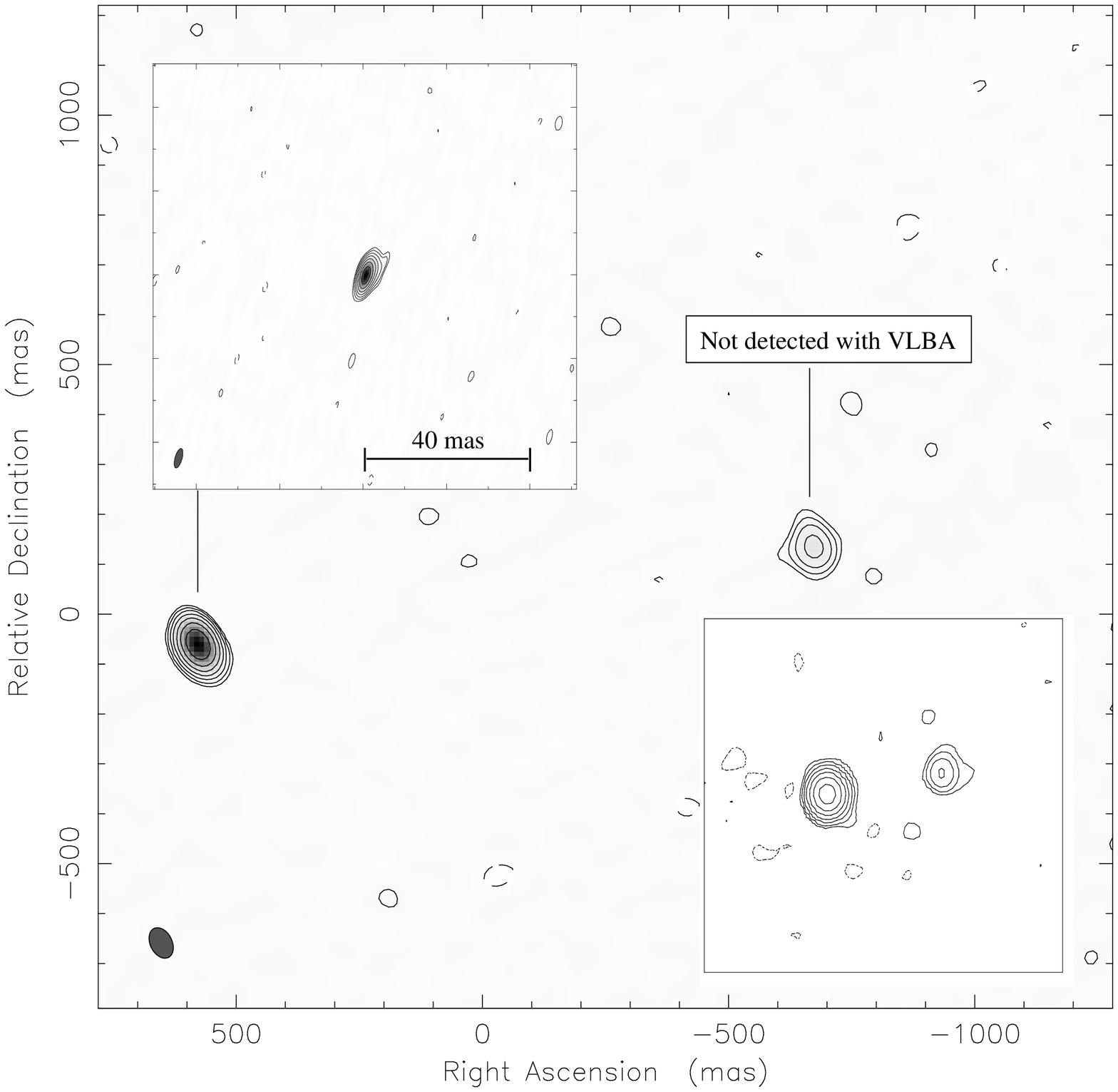}}
\end{picture}
\caption{MERLIN 5~GHz map of the rejected lens candidate
J0722+195. The VLA 8.4~GHz map is shown as an inset in the bottom
right hand corner. Also shown is the VLBA 5~GHz map of the stronger,
eastern, component.  In the VLBA observation the weaker, western, component is
not seen.
\label{0725_merlin}}
\end{figure*}

{\bf J0722+195} Both components of this 6:1 double are seen to be compact in a
MERLIN 5~GHz map with 50~mas resolution (Figure~\ref{0725_merlin}). When observed with
the VLBA at milliarcsec resolution, however, the weaker secondary
component was not detected and thus is of much lower surface brightness
than the compact primary. The nature of this secondary component is not clear
but it is not an image of the primary.

\begin{figure*}
\centering
\setlength{\unitlength}{1in}
\begin{picture}(7.5,6.5)
\put(1.6,0.3){\includegraphics{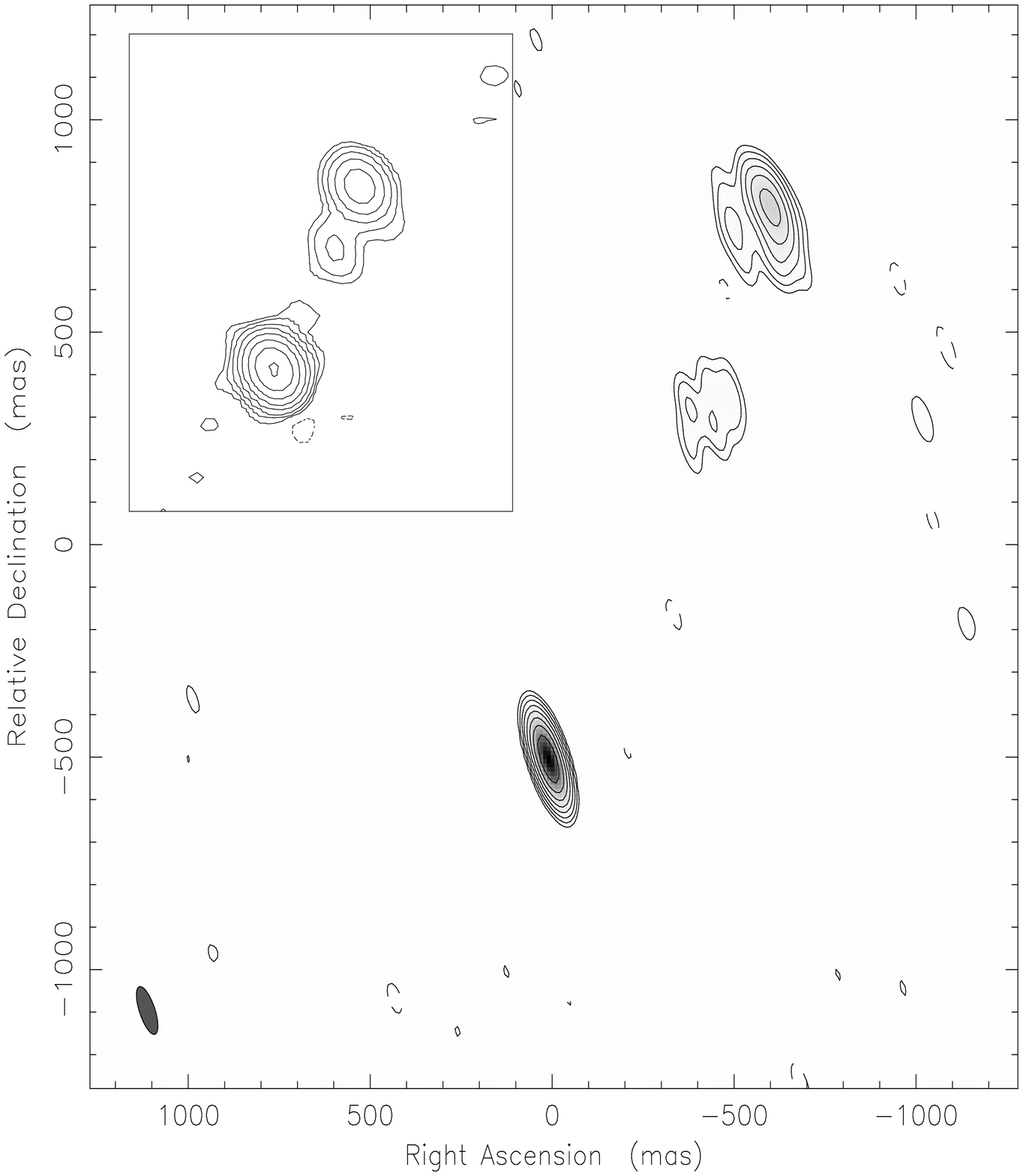}}
\end{picture}
\caption{MERLIN 5~GHz map of the rejected lens candidate J1411+006. The 8.4~GHz vla map is inset for comparison   \label{1411_merlin}}
\end{figure*}

{\bf J1411+006} In the VLA 8.4~GHz finding map (inset)  two compact
components with a flux-density ratio of 5:1 and separated by
1.4~arcsec are seen. The MERLIN 5~GHz map (Figure~\ref{1411_merlin}) shows the secondary
to be of lower surface brightness than the primary and, in addition, a
new component is detected between the two main ones suggesting
that the structure consists of a core and knots in a faint underlying  jet.

\subsubsection{Examples of difficult candidates}

There were just three candidates after extensive observations with
MERLIN, VLBA and HST which could not be ruled out as, or confirmed as,
lensed systems with one hundred percent certainty.

\begin{figure*}
\centerline{
\psfig{figure=0305_merlin.vps,width=7cm,angle=0}
\psfig{figure=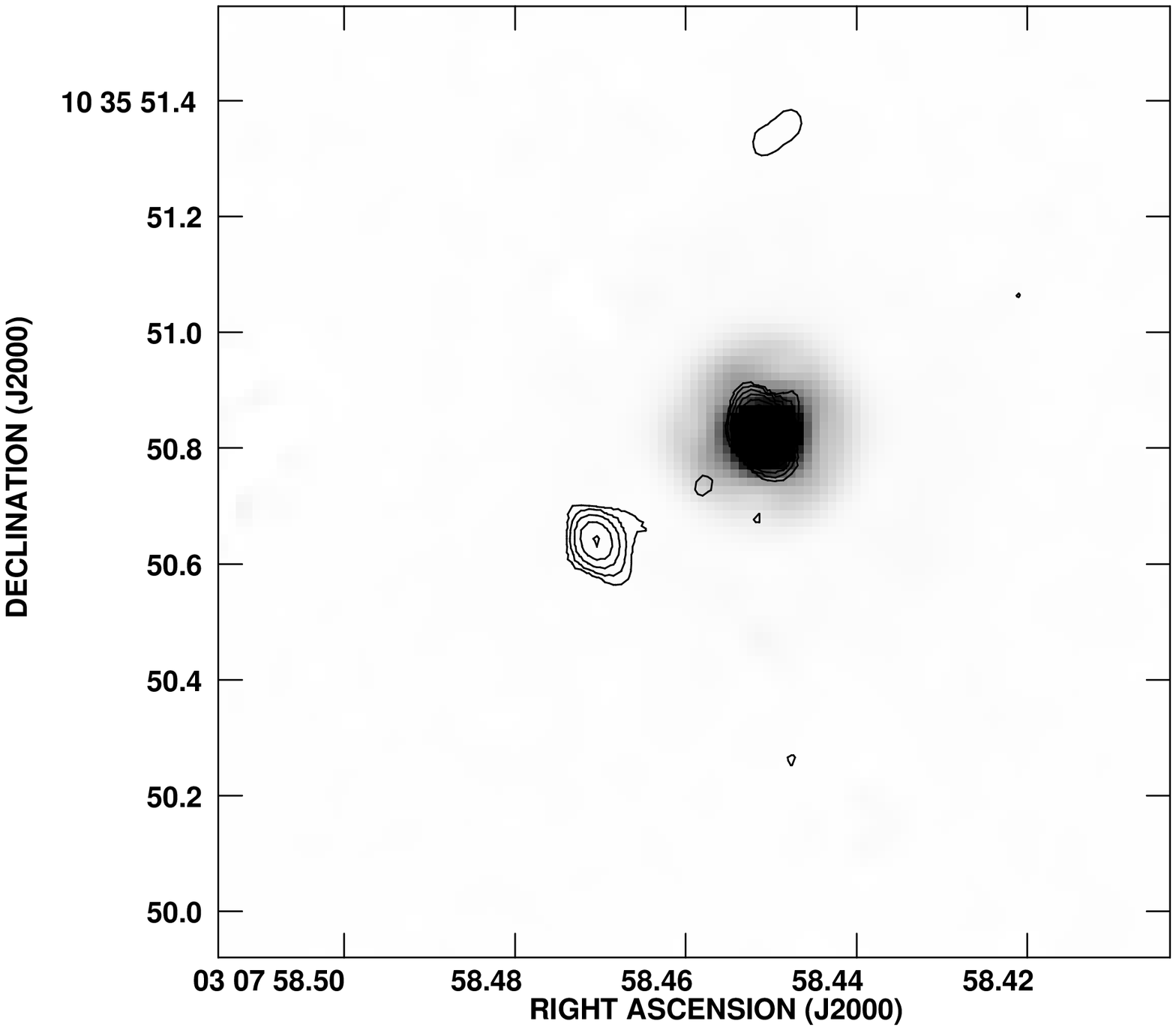,width=7.5cm,angle=0}}
\centerline{
\psfig{figure=0305_1.7nat.vps,width=7cm,angle=0}}
\caption{Radio and infrared images of J0307+106. Top row, left: MERLIN
5~GHz map of J0307+106 showing the two compact components. Top row,
right: The same MERLIN map overlayed on to the HST H-Band image of the
field obtained with NICMOS. There is no evidence of extension in the
NICMOS image. It should be noted that the accuracy of the registration
of the NICMOS and the MERLIN pictures is $\sim$0.5~arcsec so it is not
possible identify which, if any, of the radio components is coincident
with the NICMOS object. Below: A naturally weighted 1.7~GHz VLBA map of
J0307+106 restored with a 4~mas circular beam.}
\label{0307maps}
\end{figure*}

{\bf J0307+106} The MERLIN 5~GHz map (Figure~\ref{0307maps}) shows
two compact radio components separated by 345~mas with a flux-density
ratio of 6:1. Additional MERLIN and VLBA 1.7 and 5~GHz maps have been
made and HST NICMOS (H) and WFPC2 (V and I) pictures have been
obtained. Both radio images are dominated by compact emission even
with the 3~mas resolution of the 5~GHz VLBA maps. Both components also
have flat radio spectra over the frequency range 1.7 to 8.4~GHz. A
relatively bright (H=16.7) compact object is detected on the NICMOS
picture (Figure~\ref{0307maps}). Neither the NICMOS nor the WFPC data
show any hint of the multiple components expected if there were both a
lens and lensed images present. The strongest evidence against the
hypothesis that J0307+106 is a lens system comes from a long track
VLBA observation at 1.7~GHz. The resulting map
(Figure~\ref{0307maps}) shows that the weaker component has a
jet-like feature pointing towards the stronger component. If it were a
lens system this jet should be easily visible in the higher
magnification northern image. However, it is surprising, in the non-lensing
hypothesis, that both components are so compact. The only way
that the lensing hypothesis could remain viable would be if the
extended ``jet'' emission were associated with the lensing
galaxy rather than the background, lensed, radio source. Although the VLBA image makes this seems highly unlikely, we cannot rule out
lensing for this source with one hundred percent certainty.

{\bf J0831+524} has been extensively followed up with both radio and
optical observations and discussed in detail by Koopmans et
al. (2000a).  In this case MERLIN and VLBA maps show two compact
components with a separation of 2.85~arcsec. In all other cases (22
out of 23 in CLASS) the existence of two or more such components has
been a reliable indicator of lensing. However, though both components
have sub-mas structure, there is a hint that the weaker component is
the more extended. Moreover, the radio spectral indices of the two
components are significantly different. Thus the radio evidence for
lensing is not entirely convincing. Optical/infrared emission is
detected from the region of both radio components but with a very
different flux-density ratio, 16:1, compared with the 2.8:1 observed
at 8.4~GHz. We would not expect much
extinction of images in a relatively symmetrical lens system with an
image separation of 2.85~arcsec since both light paths will pass
$\sim$10~kpc from the centre of the galaxy. The infrared images of the
weaker component suggests that it is somewhat extended and there is no
evidence for a lensing galaxy at the expected position visible in a
one orbit H-band NICMOS picture. The accumulated evidence therefore
suggests that J0831+524 is not a lens system and what we are seeing is
two independent (but possibly related) radio sources. There is a
redshift of 2.064 for the brighter optical counterpart but none for the
weaker (Koopmans et al., 2000a). If this system has nothing to do with
lensing, it is, nevertheless, highly unusual in presenting two sub-mas
components with a projected separation of galactic dimensions.

\begin{figure*}
\centering
\setlength{\unitlength}{1in}
\begin{picture}(7.5,6.5)
\put(1.6,0.3){\includegraphics{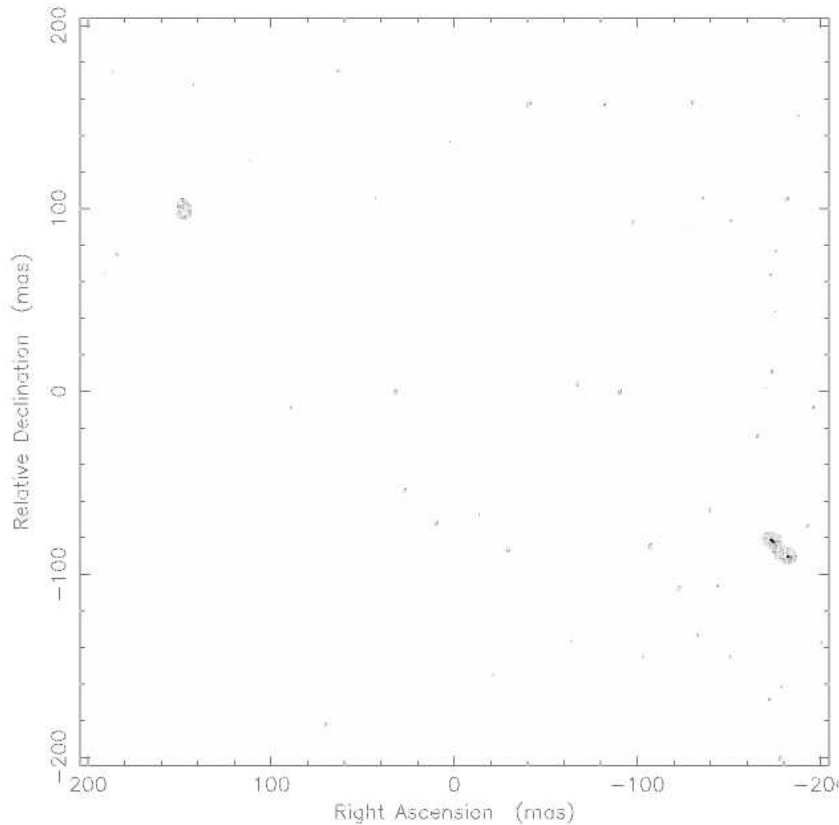}} 
\end{picture}
\caption{A naturally weighted 5~GHz VLBA map of J0935+073 restored with a 3~mas circular beam.\label{0932_vlba}}
\end{figure*}

{\bf J0935+073} has a component separation of 370~mas but a
flux-density ratio 22.6:1; hence it would not be part of the
statistically well-defined {\it lens} sample, even if it were a lens
system.  It has, nevertheless, been extensively followed up with radio
and optical observations (King et al., 1999). Both components are
detected in a VLBA map at 5~GHz (Figure~\ref{0932_vlba}). This map
shows that the strong primary component resembles a CSO with its axis
pointing approximately towards the weak secondary. Optically there is
an emission line galaxy, with a redshift of 0.28, at the position of
the radio components (but the astrometry is not good enough to
identify which component is coincident with the optical
emission). There is also a detection of a 260~mJy IRAS 60$\mu$ source
near the radio position.

The primary and secondary components are almost certainly related
because of their proximity and the fact that the primary points
towards the secondary. In the lensing scenario, however, one would not
expect the radial stretching seen in the primary. Also the presence of
a low-redshift AGN is naturally explained if it is the origin of the
radio emission but having an AGN as the lens would have to be attributed to
coincidence in the lensing hypothesis.  The CSO-like radio structure
of the primary fits the scenario of the whole system being associated
with the AGN, though the secondary radio component does not have an
obvious interpretation in this picture. While we think gravitational
lensing is an unlikely explanation for J0935+073 it still cannot be
completely ruled out.

\begin{table*}
\begin{tabular}{ l l l l l l l l}
{\bf IAU name}&\multicolumn{1}{c}{\bf Coordinates}&{\bf Follow-up}&\multicolumn{4}{c}{\bf Reasons for rejection}&{\bf Notes}\\
\multicolumn{1}{c}{\bf J2000}&\multicolumn{1}{c}{\bf J2000}&{\bf
telescope}&{\bf s.b.}&{\bf s.i.}&{\bf p.}&{\bf m.}\\ \hline

J0000+393  & 00 00 41.5259  +39 18 04.172   &MERLIN&X&&X&&\\

J0009+400  & 00 09 04.1723  +40 01 46.677 &MERLIN&X&&&&\\

J0010+490$^{*}$  & 00 10 02.3130  +49 01 58.584 &MERLIN,VLBA&X&&&&\\
 
J0026+351  & 00 26 41.7238 +35 08 42.24&MERLIN&&X&&&Flux ratio $\geq$10:1\\

J0028+17   & 00 28 30.0607  +17 07 44.715 &MERLIN&X&&&X&\\

J0058+141  & 00 58 17.8040  +14 10 48.173 &MERLIN&&&&X&\\

J0100+511  & 00 59 59.4186  +51 07 31.154 &MERLIN,VLBA&X&&&X&$\leq$300~mas\\

J0101+115  & 01 01 28.2756  +11 35 31.495 &MERLIN&&&&X&$\leq$300~mas\\

J0108+198  & 01 09 00.8921  +19 49 55.855 &MERLIN,VLBA&X&&&&\\

J0111+134$^{*}$  & 01 11 36.5675  +13 24 37.555 &MERLIN&X&&&&\\

J0112+379  & 01 12 00.2598  +37 59 31.559 &MERLIN,VLBA&X&&&&\\

J0138+293$^{*}$ & 01 38 35.3186 +29 22 04.493 &MERLIN&X&X&&&\\

J0152+338$^{*}$ & 01 52 34.5718 +33 50 33.232&MERLIN&&X&&X&\\

J0210+211 & 02 10 59.9977  +21 10 54.641 &MERLIN&X&&&& $\leq$300~mas\\

J0212+106 & 02 12 42.9722  +10 41 43.369 &MERLIN&X&&&X&\\

J0221+359$^{*}$ & 02 21 05.4729 +35 56 13.819&MERLIN,VLBA&&&&&  B0218+357:LENS\\

J0238+155 & 02 38 19.8898  +15 33 22.530 &MERLIN&X&&&&\\

J0255+026$^\dagger$ & 02 55 31.6461  +02 40 22.654 &VLA&X&X&&&\\

J0307+106 & 03 07 58.4529  +10 35 50.835 &MERLIN,VLBA &&&&& see Section 4.1.2\\

J0319+310$^{*}$ & 03 19 20.7206 +31 02 11.568 &MERLIN&X&&&&\\

J0422+157 & 04 22 53.6481  +15 47 33.688 &MERLIN,VLBA&X&&&&\\

J0430+063 & 04 30 12.5330  +06 20 33.373 &MERLIN&X&&&&\\

J0432+639 & 04 32 53.9461  +63 56 40.131 &VLA&&&&X&\\

J0448+124 & 04 48 21.9905  +12 27 55.388 &MERLIN,VLBA,VLA&&&&&  B0445+123:LENS\\

J0448+098 &  04 48 21.7383  +09 50 51.480 &MERLIN&&&&X&\\

J0458+201$^{*}$ & 04 58 29.8665 +20 11 36.115&MERLIN&X&&&&\\

J0537+645 & 05 37 52.3575  +64 34 11.649  &MERLIN&X&&&X&\\

J0552+726$^{*}$ & 05 52 52.9690 +72 40 44.995&MERLIN&X&X&&&\\

J0558+534$^{*}$ & 05 58 11.8134 +53 28 17.664&MERLIN&X&&&&\\

J0605+627 & 06 05 04.4474  +62 46 53.997 &MERLIN&X&&&&\\ 

J0635+519 & 06 35 12.3120  +51 57 01.788 &MERLIN,VLBA&&&&& B0631+519:LENS\\

J0641+356$^{*}$ & 06 41 35.8456 +35 39 57.728&MERLIN&X&&&&\\

J0644+465 & 06 44 35.3632  +46 31 17.083 &MERLIN&X&&&&\\

J0656+324 & 06 56 33.6131  +32 28 30.091 &MERLIN&X&&&&\\

J0704+450 & 07 04 50.9609  +45 02 41.623 &MERLIN&X&&&&\\

J0706+531 & 07 06 07.3296  +53 09 55.157 &MERLIN,VLBA&X&&&X&\\

J0716+471 & 07 16 03.5799  +47 08 50.063 &MERLIN,VLBA&&&&&  B0712+472 LENS\\

J0725+195 & 07 25 33.0763   +19 32 14.799 &MERLIN,VLBA&X&&&&\\

J0732+458 & 07 32 19.9471  +45 50 40.150 &MERLIN&X&&&&\\

J0734+165 & 07 33 59.6960  +16 31 05.760 &MERLIN&X&&&&\\

J0749+578$^{*}$ & 07 49 56.9552 +57 50 15.204&MERLIN&X&&&&\\

J0753+094$^{*}$ & 07 53 51.9407 +09 24 19.760&MERLIN&X&&&&\\

J0756+449 & 07 56 47.5367  +44 58 55.112 &MERLIN&X&&&X&\\

J0802+529 & 08 02 43.9915  +52 55 44.046 &MERLIN,VLA&X&X&&&\\

J0803+450 & 08 03 40.5128  +45 00 56.937 &MERLIN&X&&&&\\

J0812+406 & 08 12 03.0283   +40 41 08.205   &MERLIN&X&&&& Flux density ratio $\geq$10:1\\

J0824+392$^{*}$ & 08 24 55.4803 +39 16 41.982 &MERLIN&X&X&X&&\\

J0828+569 & 08 28 16.3028  +56 59 30.062 &MERLIN&X&&&&\\

J0831+524 & 08 31 05.4790  +52 25 21.157   &MERLIN,VLBA&&X&&&see Section 4.1.2\\

J0832+424$^{*}$ & 08 32 48.4047 +42 24 59.131 &MERLIN&X&X&X&&\\

J0834+440 & 08 34 58.2033  +44 03 38.147 &MERLIN&X&&&&Flux density ratio $\geq$10:1\\

J0840+193 & 08 40 44.7018  +19 19 11.608 &MERLIN&X&&&&\\

J0852+052 & 08 52 53.5725  +05 15 15.654 &MERLIN,VLBA&&&&&  B0850+054:LENS\\

J0855+487 & 08 55 45.3300  +48 42 25.285 &MERLIN&X&&&\\

J0856+046 & 08 56 17.9800  +04 38 41.550 &MERLIN&X&&&X&\\

J0903+468$^{*}$ & 09 03 03.9905 +46 51 04.133&MERLIN&X&&&&\\

J0910+196 & 09 10 26.2070  +19 36 43.954 &MERLIN&X&&&X&\\

J0912+414 &  09 12 11.6157   +41 26 09.355   &MERLIN&X&&&&\\

J0921+716$^{*}$ & 09 21 23.9475 +71 36 12.311 &MERLIN&X&X&X&&\\

J0930+449 & 09 30 14.3955  +44 57 26.425 &MERLIN&X&&&&\\

J0935+073 & 09 35 01.0733 +07 19 18.660 &MERLIN,VLBA&X&&&&see Section 4.1.2\\

J0936+264 & 09 36 14.2430 +26 24 08.100  &MERLIN&X&&&&\\

J0949+662$^{*}$ & 09 49 12.1520 +66 14 59.671&MERLIN&X&&&&\\

\end{tabular}
\end{table*}
\begin{table*}

\begin{tabular}{l l l l l l l l}
{\bf IAU name}&\multicolumn{1}{c}{\bf Coordinates}&{\bf Follow-up}&\multicolumn{4}{c}{\bf Reasons for rejection}&{\bf Notes}\\
\multicolumn{1}{c}{\bf J2000}&\multicolumn{1}{c}{\bf J2000}&{\bf telescope}&{\bf s.b.}&{\bf s.i.}&{\bf p.}&{\bf m.}\\
\hline

J0958+298$^\dagger$ & 09 58 58.9469  +29 48 04.179 &Optical&&&&&Independent quasars (Leh\'{a}r et al)\\

J1000+278$^\dagger$ & 10 00 29.1481  +27 52 12.168 &Optical&&&&&Independent quasars (Phillips et al)\\

J1012+331 & 10 12 11.4526  +33 09 26.414 &VLA&&X&X&&\\


J1034+594 & 10 34 34.2393  +59 24 45.846  &MERLIN&X&&&X&\\

J1040+036 & 10 40 37.4577  +03 41 59.774 &MERLIN&&&&X&\\

J1124+167 & 11 25 00.6294  +16 44 08.145 &MERLIN&&&&X&\\


J1131+517 & 11 31 16.4545  +51 46 34.276 &MERLIN,VLA&X&&&X&\\

J1132+604 &    11 32 58.7470   +60 29 57.151     &MERLIN&X&&&&\\

J1144+221 & 11 44 17.8365  +22 07 53.627  &MERLIN&X&&&&\\

J1155+196 & 11 55 18.3002  +19 39 42.234 &MERLIN,VLBA&&&&&  B1152+199:LENS\\

J1213+131$^{*}$ & 12 13 32.1554 +13 07 20.556&MERLIN&X&&&&\\

J1217+490 & 12 17 49.5808  +49 02 04.571 &MERLIN&X&&&X&\\

J1236+393$^{*}$ & 12 36 51.4519 +39 20 27.850 &MERLIN&X&X&X&X&\\

J1240+262$^\dagger$ & 12 40 02.1435  +26 17 20.687 &VLA&&X&&&\\

J1246+001 & 12 46 02.8349  +00 07 55.144 &MERLIN&X&&&&\\

J1247+214 & 12 47 53.8655  +21 27 58.114 &MERLIN,VLBA&X&&&X& Component separation $\leq$300~mas\\

J1252+191 & 12 52 27.8405  +19 10 38.182 &MERLIN&X&&&&\\

J1255+614 & 12 55 45.0142  +61 24 50.872 &MERLIN&X&&&X&\\

J1312+735 &   13 12 48.2953   +73 35 40.634  &MERLIN&X&&&&\\

J1314+087 & 13 14 03.1904  +08 42 09.056 &MERLIN&X&&&&\\

J1317+344$^{*}$ & 13 17 36.4935 +34 25 15.923 &MERLIN&X&X&&X&\\

J1326+607$^\dagger$ & 13 26 43.9112  +60 42 21.591 &VLA&&X&&&\\

J1329+108 & 13 29 01.4188  +10 53 04.799 &MERLIN&X&&&&\\

J1330+138 & 13 30 54.1248  +13 51 03.064 &MERLIN&X&&&X&\\

J1401+152 & 14 01 35.5502  +15 13 25.638 &MERLIN,VLBA&&&&& B1359+154:LENS\\

J1402+037 & 14 02 24.8410  +03 42 26.588 &MERLIN&X&&&&\\

J1411+006 & 14 11 07.8383  +00 36 07.200 &MERLIN&X&&&X&\\

J1423+246 & 14 23 35.4922  +24 36 20.566 &MERLIN&X&&&X&\\

J1424+229$^{*}$ & 14 24 38.0544 +22 56 00.036&MERLIN,VLBA&&&&& B1422+231:LENS\\

J1429+541$^{*}$ & 14 29 21.8824 +54 06 11.215 &MERLIN&X&X&&&\\

J1431+731 & 14 31 56.2148  +73 10 40.887 &MERLIN&X&&&&\\

J1432+363 &  14 32 39.8290   +36 18 07.946   &MERLIN&X&&&&Flux density ratio $\geq$10:1\\

J1440+059$^{*}$ & 14 40 17.9825 +05 56 34.090 &MERLIN&X&&&&\\

J1442+526 & 14 42 19.4588  +52 36 21.706 &MERLIN&X&&&X&\\ 

J1446+144 & 14 46 42.3720  +14 28 01.458 &VLA&X&&&&\\

J1452+099 & 14 52 25.5265  +09 55 46.607 &MERLIN,VLBA&X&&&&\\

J1501+563 & 15 01 24.6325  +56 19 49.655 &MERLIN&X&&&X&\\

J1514+509 & 15 14 01.4944  +50 54 29.847 &MERLIN&&X&&X&\\

J1521+312$^{*}$ & 15 21 01.2836 +31 15 37.961  &MERLIN&X&&&&Flux density ratio $\geq$10:1\\

J1528+058 & 15 28 44.5754  +05 52 17.402 &VLBA&X&&&&\\

J1540+147$^{*}$ & 15 40 49.4620 +14 47 45.959 &MERLIN&X&&&&\\

J1545+529 & 15 45 04.9061  +52 59 25.475 &MERLIN&X&&&&\\

J1546+082 & 15 46 00.6754  +08 15 03.076 &MERLIN,VLBA&X&&&&\\

J1546+448 & 15 46 04.4211  +44 49 10.529 &MERLIN&X&&&&\\

J1609+655 & 16 09 13.9581  +65 32 28.975 &MERLIN,VLBA&&&&& B1608+656:LENS\\

J1621+099 & 16 21 39.3805  +09 59 20.538 &MERLIN&X&&&& \\

J1625+408 & 16 25 10.3249  +40 53 34.339 &MERLIN&X&&&&\\

J1631+449 & 16 31 32.3744  +44 58 49.303 &MERLIN,VLBA&X&&&&\\

J1641+512 & 16 41 55.7387  +51 15 46.884 &MERLIN&X&&&X&Flux density ratio $\geq$10:1\\

J1657+260 &  16 57 14.2163  +26 00 28.949 &MERLIN&&&&&Component separation $\leq$300~mas\\

J1702+552 & 17 02 34.5570  +55 11 12.432  &MERLIN,VLA&&&&X&\\

J1707+232 & 17 07 25.4544  +23 12 21.584 &MERLIN,VLBA&X&&&&\\

J1708+110 & 17 08 25.9547  +11 04 53.091 &MERLIN&X&&&&\\

J1722+561 & 17 22 58.0083  +56 11 22.320 &MERLIN&X&&&X&\\

J1724+045 & 17 24 52.0601  +04 35 00.390 &MERLIN&X&X&&&\\

J1724+453 & 17 24 35.4501  +45 20 14.909 &MERLIN&&&&X&\\

J1726+399$^{*}$ & 17 26 32.6614 +39 57 02.178 &MERLIN&X&X&X&&\\

J1750+153 & 17 50 05.0697  +15 18 42.871 &MERLIN&X&&&&\\

\end{tabular}
\end{table*}
\begin{table*}

\begin{tabular}{l l l l l l l l}
{\bf IAU name}&\multicolumn{1}{c}{\bf Coordinates}&{\bf Follow-up}&\multicolumn{4}{c}{\bf Reasons for rejection}&{\bf Notes}\\
\multicolumn{1}{c}{\bf J2000}&\multicolumn{1}{c}{\bf J2000}&{\bf telescope}&{\bf s.b.}&{\bf s.i.}&{\bf p.}&{\bf m.}\\
\hline

J1802+268$^{*}$ & 18 02 32.3014 +26 53 28.963 &MERLIN&X&&&&\\

J1819+307 & 18 19 01.4108  +30 42 23.871 &MERLIN&X&&&&\\

J1821+359 & 18 21 40.0122  +35 57 56.445 &MERLIN&X&&&&\\

J1835+490 & 18 35 21.8102  +49 04 43.612 &MERLIN,VLA&X&&X&&\\

J1836+183 & 18 36 15.6844  +18 23 49.768 &MERLIN,VLBA&X&&&&\\

J1928+421 &  19 28 21.7712  +42 06 21.566    &MERLIN&&&&X&\\

J1934+504 & 19 34 30.8954  +50 25 23.213 &MERLIN,VLBA&&&&& B1933+503:LENS\\

J1937+648 & 19 37 12.0854  +64 52 20.881 &MERLIN&X&&&&\\


J1947+678 & 19 47 36.2599  +67 50 16.928 &MERLIN,VLBA&X&&&X&\\

J2030+086 &  20 30 04.3844   +08 39 37.536   &MERLIN&&&&X&\\

J2047+267 & 20 47 20.2885  +26 44 02.699 &MERLIN,VLBA&&&&& B2045+265:LENS\\

J2049+127 & 20 49 14.4502  +12 46 42.049 &MERLIN,VLBA&X&&&&\\

J2049+073$^\dagger$ & 20 49 53.8362  +07 18 59.440 &MERLIN&&&X&&\\


J2116+024$^{*}$ &  21 16 50.7461 +02 25 46.462 &MERLIN,VLBA&&&&& B2114+022:LENS\\

J2125+328 & 21 25 18.8461  +32 52 03.907 &MERLIN&X&&&&\\

J2127+103 & 21 27 25.2300  +10 18 45.700 &MERLIN&&&&X&\\

J2139+190 & 21 39 49.5744  +19 04 34.068 &MERLIN&X&&&X&\\

J2200+105$^{*}$ & 22 00 07.9332 +10 30 07.798 &MERLIN&X&&&&\\

J2209+359$^{*}$ & 22 09 45.3360 +35 56 01.017 &MERLIN&&X&X&&\\

J2220+264 &    22 20 22.0922   +26 28 04.490   &MERLIN&X&&&&\\

J2241+288 & 22 41 14.2516  +28 52 20.119 &MERLIN&&&&X&\\

J2250+461 & 22 50 55.3802  +46 06 34.478 &MERLIN,VLA&&&X&&\\

J2255+434$^\dagger$ & 22 55 11.2587  +43 28 22.359 &MERLIN&X&X&&&\\

J2321+454 & 23 21 09.0308  +45 25 42.439 &VLA&&X&&&\\

J2321+054 &   23 21 40.8044  +05 27 37.210   &MERLIN,VLBA&&&&& B2319+051:LENS\\ 

J2344+278$^{*}$ & 23 44 37.0557 +27 48 35.422 &MERLIN&X&X&X&&\\

J2355+228 & 23 55 27.4704  +22 53 18.093 &MERLIN&X&&&&\\

J2358+393$^{*}$ & 23 58 59.8442 +39 22 28.322 &MERLIN&X&X&&&\\

\end{tabular}
\caption{149 candidates for gravitationally lensed sources selected
from the CLASS complete sample: the first and second columns show the
J2000 IAU names and J2000 coordinates respectively. Sources marked
with a $^*$ are JVAS objects and have already been reported by King et
al.  (1999). Sources marked with a $^\dagger$ have been reported by
Phillips et al. (2001) and have component separations in the range 6
to 15~arcsec.  In column 3 the instruments used for the follow-up observations are
listed. Columns 4 through 7 indicate the reasons behind the rejection
of candidates after the follow-up observations; the surface
brightnesses (s.b.), spectral indices (s.i.) and polarisations (p.) of
each candidate's components are compared in columns 4 through 6
respectively, and marked with a X where they are sufficiently
dissimilar to rule out the candidate as a lens system. An X in column
7 shows that there was a morphological (m.)  reason for ruling out the
candidate. Notes on individual objects are given in column
8.\label{cand}}
\end{table*}



\section{Results of follow-up observations: Gravitational lens systems}

\begin{figure*}
\centering
\setlength{\unitlength}{1in}
\begin{picture}(7.5,8.5)
\put(0.5,-1.0){\includegraphics{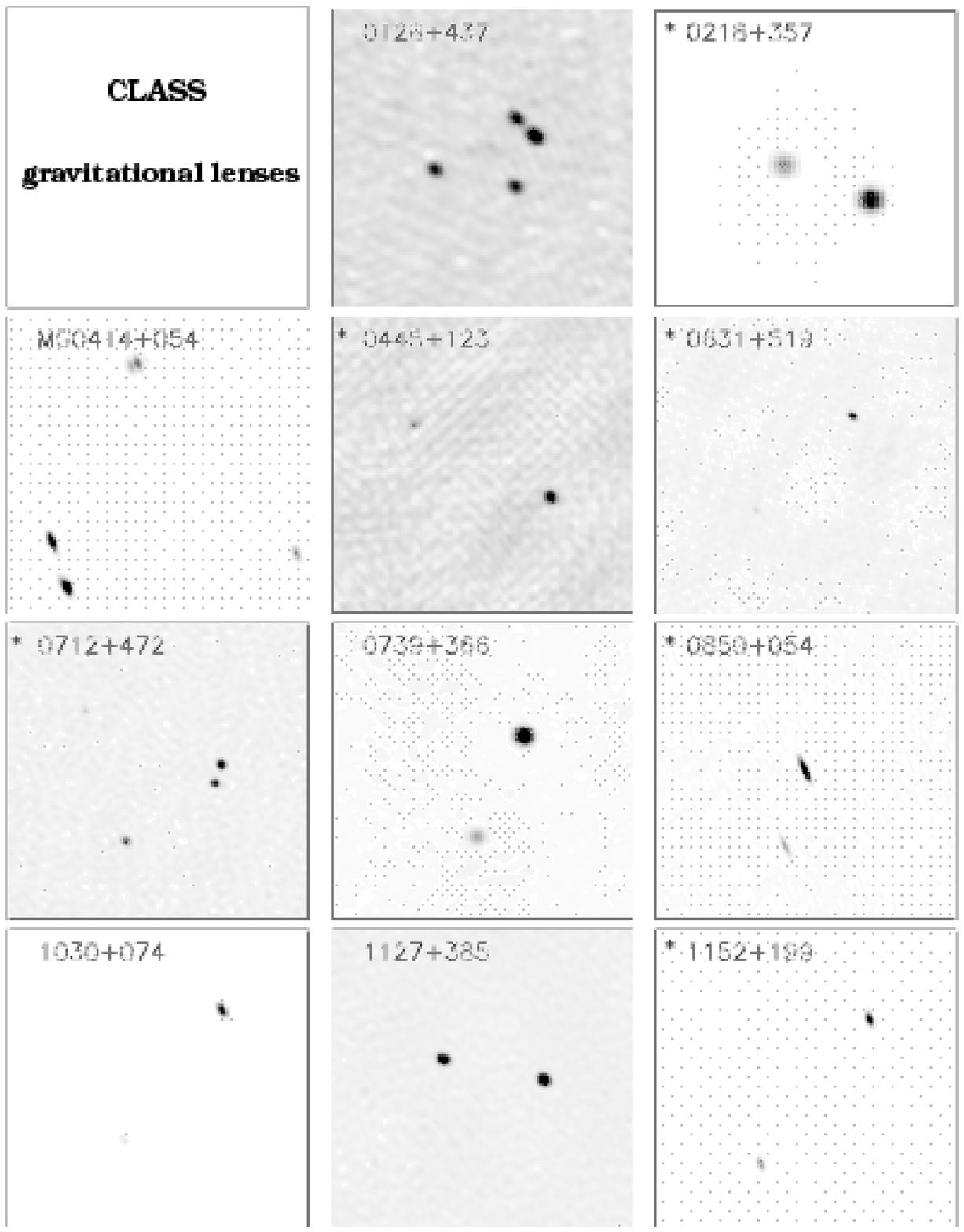}}
\end{picture}
\caption{MERLIN 5~GHz maps of CLASS lens systems. Members of the statistically weel-defined sample are marked with an asterisk \label{merlin1}}
\end{figure*}

\begin{figure*}
\centering
\setlength{\unitlength}{1in}
\begin{picture}(7.5,8.5)
\put(0.5,-2.0){\includegraphics{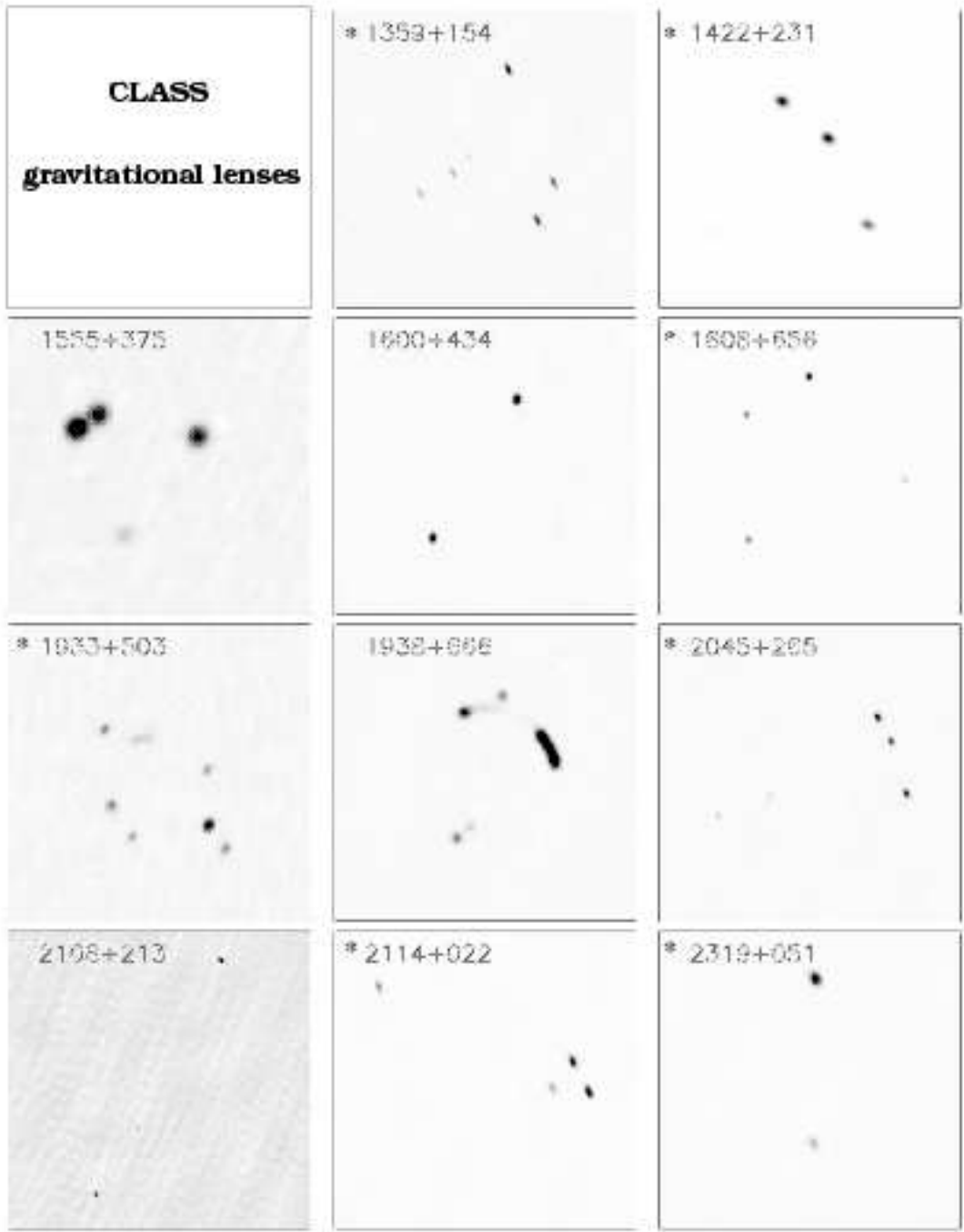}}
\end{picture}
\caption{MERLIN 5~GHz maps of CLASS lens systems. \label{merlin2}}
\end{figure*}

In this section we summarise the properties of the 13 gravitational
lens systems contained in the statistically well-defined CLASS lens
sample (i.e. satisfying the criteria set out in Section 2) and the
nine additional lens systems that have been found in CLASS but which
do not satisfy the criteria for inclusion in the statistically
well-defined sample. In Table \ref{complete} we list the well-defined
sample systems, their maximum image separation ($\Delta\theta$), image
multiplicity (N$_{\rm im}$), and the redshifts of the background
lensed source (z$_{s}$) and lensing galaxy (z$_{l}$) where known. In
Table \ref{additional} we list the nine other lens systems. Some of
these are not members of the complete sample (i.e. as defined by NVSS
and GB6) while others were followed up simply because they ``looked
promising'' even though they did not meet one or more of the
quantitative lensing criteria set out in Section 3. Redshifts for 16
of the lensing galaxies and 11 of the lensed objects are available
(see Table~\ref{complete} and Table~\ref{additional}).  MERLIN 5~GHz maps
of each of the 22 lens systems are shown in Figure~\ref{merlin1} and
Figure~\ref{merlin2}. Each individual system is described briefly
below.

\begin{table*}
\caption{Gravitational lenses in the CLASS statistically well-defined
sample. Column 1, B1950.0 name; Columns 2 \& 3, J2000.0 coordinates; Column 4, the origin of the system; Columns 5 \& 6, source and lens redshifts respectively; Column 7, the maximum  image separation given in arcsec; Column 8 the ratio of the flux densities of the images (for the doubles only); Column 9, the number of images. The redshift
references are; $^{1}$Lawrence (1996), $^{2}$Browne et al. (1993),
$^{3}$Jackson et al. (1998), $^{4}$Fassnacht \& Cohen
(1998),$^{5}$Myers et al. (1999), $^{6}$Patnaik et al. (1992b),
$^{7}$Kundi{\'c} et al. (1997), $^{8}$Fassnacht et al. (1996), $^{9}$Myers et
al. (1995), $^{10}$Norbury et al. (2002), $^{11}$Sykes et al. (1998),
$^{12}$Fassnacht et al. (1999a), $^{13}$Augusto et al. (2001), $^{14}$Lubin et al. (2000), $^{15}$Argo et al. (2002), $^{16}$Biggs et al. (2003).
\label{complete}}
\begin{tabular}{lllllllll}
\hline 
1  &  2  &  3   &  4 & 5 & 6      & 7 & 8 & 9  \\
\hline
B0218+357  & 02 21 05.4729 & +35 56 13.819 & JVAS     &  0.96$^{1}$  & 0.68$^{2}$  & 0.334 & 3.8  & 2  \\
B0445+123  & 04 48 21.9905 & +12 27 55.388 & CLASS    &        &  0.557$^{15}$     & 1.33  & 7.3  & 2  \\
B0631+519  & 06 35 12.3120 & +51 57 01.788 & CLASS    &        &       & 1.16  & 6.6  & 2  \\
B0712+472  & 07 16 03.5799 & +47 08 50.063 & CLASS    &  1.34$^{3}$  & 0.41$^{4}$  & 1.27  &  -   & 4    \\
B0850+054  & 08 52 53.5725 & +05 15 15.654 & CLASS    &        &  0.59$^{16}$     & 0.68  & 6.9  & 2  \\
B1152+199  & 11 55 18.3002 & +19 39 42.234 & CLASS    &  1.013$^{5}$ & 0.435$^{5}$ & 1.56   & 3.0  & 2  \\
B1359+154  & 14 01 35.5502 & +15 13 25.638 & CLASS    &  3.214$^{5}$ &  -    & 1.65  &  -   & 6    \\
B1422+231  & 14 24 38.0544 & +22 56 00.036 & JVAS     &  3.62$^{6}$  & 0.34$^{7}$  & 1.28  &  -   & 4    \\
B1608+656  & 16 09 13.9581 & +65 32 28.975 & CLASS    &  1.39$^{8}$  & 0.64$^{9}$  & 2.08  &  -   & 4    \\
B1933+503  & 19 34 30.8954 & +50 25 23.213 & CLASS    &  2.62$^{10}$  & 0.755$^{11}$ & 1.17  &  -   & 4    \\
B2045+265  & 20 47 20.2885 & +26 44 02.699 & CLASS    &  1.28$^{12}$  & 0.867$^{12}$  & 1.86  &  -   & 4    \\
B2114+022  & 21 16 50.7461 & +02 25 46.462 & JVAS     &  -     &0.32/0.59$^{13}$ & 2.57& 3.0 & 2? \\
B2319+051  & 23 21 40.8044 & +05 27 37.210 & CLASS    &  -     & 0.624/0.588$^{14}$ & 1.36  & 5.7  & 2  \\
\hline
\end{tabular}
\end{table*}

\begin{table*}
\caption{Additional gravitational lens systems found in CLASS. Column 1, B1950.0 name; Columns 2 \& 3, J2000.0 coordinates; Column 4, the origin of the system; Columns 5 \& 6, source and lens redshifts respectively; Column 7, the maximum  image separation given in arcsec; Column 8 the ratio of the flux densities of the images (for the doubles only); Column 9, the number of images; Column 10, the
reason why each lens does not meet the criteria to be included in the
statistically well-defined sample: 1
-- not a member of the CLASS complete sample, 2 -- the flux density
ratio of the images is $\geq$10:1, 3 -- the object is only recognized
as being lensed by virtue of its extended radio emission, 4 -- the sum
of the image flux densities at 8.4~GHz is $\leq$20~mJy. The image
separation is given in arcsec. References to redshifts are
:$^{1}$Lawrence, Cohen \& Oke (1995), $^{2}$Tonry and Kochanek (1999),
$^{3}$Fassnacht \& Cohen (1998), $^{4}$Tonry and Kochanek (2000),
$^{5}$Koopmans et al., in preparation (2002).\label{additional}}
\begin{tabular}{llllllllll}
\hline 
1  &  2  &  3   &  4  & 5 & 6      & 7  & 8 & 9 & 10 \\
\hline
B0128+437  & 01 31 13.405 & 43 58 13.140 & CLASS    &        &       & 0.54 &  -   & 4  & 1 \\
B0414+054  & 04 14 37.770 & 05 34 42.361 & MIT/JVAS &  2.62$^{1}$  & 0.96$^{2}$  & 2.09 &  -   & 4  & 1 \\
B0739+365  & 07 42 51.169 & 36 34 43.638 & CLASS    &  -     &  -    & 0.53 & 6.4  & 2  & 1  \\
B1030+074  & 10 33 34.025 & 07 11 26.122 & JVAS     &  1.53$^{3}$  & 0.599$^{3}$ & 1.56 & 15   & 2  & 2  \\
B1127+385  & 11 30 00.099 & 38 12 03.091 & CLASS    &  -     &  -    & 0.70 & 1.2  & 2  & 1  \\
B1555+375  & 15 57 11.940 & 37 21 35.970 & CLASS    &  -     &  -    & 0.43 &  -   & 4  & 1  \\
B1600+434  & 16 01 40.500 & 43 16 44.000 & CLASS    &  1.57$^{3}$  & 0.415$^{3}$ & 1.39 & 1.2  & 2  & 1  \\
B1938+666  & 19 38 25.290 & 66 48 52.960 & JVAS     &  -     &  0.878$^{4}$    & 0.93   &  -   & 4  & 3  \\
B2108+213  & 21 10 54.140 & 21 30 59.100 & CLASS    &  -     & 0.365$^{5}$ & 4.55 & 2.3  & 3?  & 4  \\  

\hline
\end{tabular}
\end{table*}

\subsection{Descriptions of the individual lens systems}

\smallskip

\noindent{\it CLASS B0128+437}

\smallskip

Phillips et al. (2000) reported the discovery of this system. Infrared
imaging observations made with UKIRT at 2.2~$\mu$ show the lensing
galaxy and evidence for the strongest lensed image. The lensed object
has a radio spectrum which is strongly peaked at $\sim$1~GHz and hence
has a spectral index of steeper than -0.5 between 1.4 and 5~GHz,
excluding it from the CLASS complete sample. VLBI observations reveal
a wealth of complex substructure in the lensed images. Of particular
interest is that in one of the images there is no clear evidence for
the compact core which is present in all the other images. This
suggests that the image may be scatter broadened (See section 6.2 and
Biggs et al., in preparation). Barvainis \& Ivison (2002) report an
upper limit of 7~mJy to the 850$~\mu$ flux density.

\smallskip

\noindent{\it CLASS B0218+357}

\smallskip

Patnaik et al. (1993) reported that B0218+357 is a gravitationally
lensed system. Further observations made with the VLA, MERLIN and VLBI
were presented by Patnaik et al. (1993), Patnaik, Porcas \& Browne
(1995) and Biggs et al. (2002) show that the lensed images contain extensive
substructure. These flat-spectrum images are separated by 335~mas and
the stronger is surrounded by a steep-spectrum Einstein ring of
diameter 335~mas.  The radio variability of the compact images has
allowed a time delay of 10.4$\pm$0.4 days (95\% confidence) to be
measured (Biggs et al., 1999; Cohen et al., 2000). The lens is a
spiral galaxy which gives rise to many radio absorption lines in the
spectrum of the lensed source (eg. Menten \& Reid 1996, Wiklind \&
Combes 1995, Carilli, Rupen \& Yanny, 1993). Heavy extinction in the lensing
galaxy means that the stronger radio image is the fainter at optical
and infrared wavelengths (Jackson, Xanthopoulos \& Browne, 2000).

\smallskip

\noindent{\it MG 0414+0534}

\smallskip

The four--image system B0411+054 (most often referred to by its J2000
IAU name MG0414+0534) was discovered during the MIT lens survey
(Turner et al., 1989, Hewitt et al., 1992, Burke et al., 1993), and
re-discovered in the course of the JVAS. It is not part of the
statistically well-defined sample as its radio spectrum is too steep
to meet the selection criteria. The system has been detected at
850~$\mu$ and 450~$\mu$ with flux densities of 25.3~mJy and 66~mJy,
respectively, by Barvainis \& Ivison (2002).

\smallskip

\noindent{\it CLASS B0445+123}

\smallskip

This is one of the most recent CLASS lens systems to be discovered
(Argo et al.,, 2002). The two main images are compact on VLBA scales
($\sim$3~mas) but there is large-scale low surface brightness
structure seen in MERLIN and VLBA maps which means that it should be
possible to obtain useful observational constraints on the mass model. An
optical image has been obtained with the WHT which shows an extended
emission which may be a combination of the lens and lensed images. A
redshift of 0.557 has been measured from a Keck spectrum obtained by
McKean et al. (in preparation).

\smallskip

\noindent{\it CLASS B0631+519}

\smallskip

B0631+519 is also a recent discovery (York et al., in
preparation). Like B0445+123, it has much radio structure visible on
the scale of tenths of an arcsecond in addition to the two main
images. VLBA observations show that the most magnified image is
circumferentially stretched and consists of four compact
sub-components. Two galaxies, a low redshift emission-line galaxy plus
a higher redshift elliptical, have been identified spectroscopically
in the system (McKean et al., in preparation); together these galaxies
probably act as the lens.

\smallskip

\noindent{\it CLASS B0712+472}

\smallskip

Jackson et al. (1998a) reported the discovery of the 4-image system
B0712+472. The lensed object has a broad emission line suggesting that
it is a quasar but the absolute magnitude $M_{v}$ of -21.6 implies a
subluminous object. HST NICMOS images in H show the AGN host galaxy
emission stretched into arcs (Jackson, Xanthopoulos \& Browne, 2000). Barvainis \&
Ivison (2002) measure an 850$\mu$ flux density of 7.5~mJy. Fassnacht
\& Lubin (2002) have identified a foreground group of galaxies which
will contribute some external shear to the mass model.

\smallskip

\noindent{\it CLASS B0739+366}

\smallskip

The GB6 5~GHz flux density of B0739+366 is 30~mJy and therefore the
source is not in the CLASS complete sample. The discovery of the
system is reported in Marlow et al. (2001). HST WFPC2 and NICMOS
pictures show the lens and the two lensed images.  Barvainis \& Ivison
(2002) measure 850~$\mu$ and 450~$\mu$ flux densities of 28.8~mJy and
71~mJy, respectively.

\smallskip

\noindent{\it CLASS B0850+054}

\smallskip

B0850+054 is a two-image system (Biggs et al., 2003). VLBA
observations at 5~GHz show that the stronger A-image is tangentially
stretched over $\sim$20~mas and has at least 5 sub-components
suggesting it is a CSO. The weaker image is radially stretched but the
resolution of the existing VLBA map is not high enough to show the
individual sub-components. UKIRT observations in the K band (Biggs et
al., in press) show both the lens and the strongest lensed
image. Spectroscopic observation with Keck give a redshift of 0.59 for
the lensing galaxy (McKean et al., in preparation).

\smallskip

\noindent{\it CLASS B1030+074}

\smallskip

The B1030+074 system was first described by Xanthopoulos et
al. (1998). The ratio of flux densities of the two images is
$\sim$15:1 and hence it does not satisfy the criteria for it to be a
member of the statistically well-defined sample. The strongest image
has a well-defined jet but its counterpart in the de-magnified image
is hard to see. The lensing galaxy and lensed images are seen in HST
observations (Jackson, Xanthopoulos \& Browne, 2000). The lensing galaxy has a weak
companion or an asymmetric extension. Flux density monitoring with the
VLA has failed to yield a time delay (Xanthopoulos et al., in
preparation).

\smallskip

\noindent{\it CLASS B1127+385}

\smallskip

B1127+385 is not part of the CLASS complete sample because its GB6
5~GHz flux density is 29~mJy and the sample limit is 30~mJy. The
discovery of this 2-image system is reported by Koopmans et
al. (1999). In addition to the lensed images, two lensing galaxies
separated by 0.6~arcsec are seen in HST images. The system is detected
by Barvainis \& Ivison (2002) at 850~$\mu$ with a flux density of
13.9~mJy.

\smallskip

\noindent{\it CLASS B1152+199}

\smallskip

Myers et al. (1999) reported the discovery of this two-image lens
system. It has been detected as an X-ray source in the northern ROSAT
All-Sky Survey (Brinkman et al., 1997). The weaker image is highly
obscured by the lensing galaxy. HST pictures show the lensing galaxy
which has a faint companion and VLBI observations show that both radio
images consist of a compact core and a jet (Rusin et al., 2002). The
system has been observed at 850~$\mu$ by Barvainis \& Ivison and an
upper limit of 6.4~mJy put on its flux density.

\smallskip

\noindent{\it CLASS B1359+154}

\smallskip

With six lensed images of the same source, B1359+154 is unique amongst lens systems
discovered so far (Myers et al., 1999; Rusin et al., 2001a). This
configuration is the result of a particularly complex deflector, which
consists of three galaxies forming the core of a compact galaxy group
(Rusin et al., 2000; Rusin et al., 2001a).  Barvainis \& Ivison (2002)
measure 850~$\mu$ and 450~$\mu$ flux densities of 11.5~mJy and 39~mJy.
respectively.

\smallskip

\noindent{\it CLASS B1422+231}

\smallskip

The discovery of B1422+231 as a gravitational lens system was reported
by Patnaik et al. (1992b).  Kundi{\'c} et al. (1997) reported
spectroscopy, near-infrared and optical photometry showing that the
main lensing galaxy and five nearby galaxies belong to a compact group
at redshift 0.338.  Observations of the milliarcsecond polarization
structure of the lensed images have been discussed by Patnaik et
al. (1999) and a tentative time delay derived from VLA monitoring has
been determined by Patnaik \& Narasimha (2001). The system has been
detected in X-rays (Chartas, 2000; Reeves \& Turner, 2000).

\smallskip

\noindent{\it CLASS B1555+375}

\smallskip
                               
Marlow et al., (1999a) reported the discovery of this system.
B1555+375 is not part of the CLASS complete sample because its GB6
5~GHz flux density is 27~mJy and the sample limit is 30~mJy. Optical
imaging with the Keck II Telescope at R band shows a faint extended
object also seen in the HST I image. We estimate the combined emission
from the lens and background source to be R = 25 mag.  Observations at
H band with the William Herschel Telescope also detected this extended
object. The combined lens and background source magnitude was measured
to be H = 19 mag. Barvainis \& Ivison (2002) measure an upper limit of
6.5~mJy to the 850~$\mu$ flux density.

\smallskip

\noindent{\it CLASS B1600+434}

\smallskip

Jackson et al. (1995) reported the discovery of B1600+434. It is the
clearest example of lensing by a spiral galaxy, which in this case 
is seen edge-on (Jaunsen \& Hjorth, 1997; Koopmans, de Bruyn \& Jackson,
1998). A radio time delay of 47$\pm$6 days between the two images has
been measured by Koopmans et al. (2000b) and an optical delay of
51$\pm$4 days has been reported by Burud et al. (2000).  Radio microlensing has been
discovered by Koopmans \& de Bruyn (2000). The 5~GHz flux density of
the source has been declining, causing it to drop out of the CLASS
complete sample due to the change in flux density between 87GB and the
more recent GB6 survey.  Barvainis \& Ivison
(2002) measure a 850~$\mu$ flux density of 7.3~mJy.

\smallskip

\noindent{\it CLASS B1608+656} 

\smallskip

B1608+656 is a 4-image lens system (Myers et al., 1995) in which the
core of an extended radio galaxy with a post starburst spectrum
(Fassnacht et al., 1996) is multiply imaged (Snellen et al.,
1995). HST observations reveal the host galaxy of the lensed object
stretched into extensive arcs (Jackson, Nair \& Browne, 1998). The
lens seems to consist of two galaxies. Time delays between the four
radio images have been determined with VLA monitoring observations by
Fassnacht et al. (1999b,2002) (see Table~\ref{delays}). Barvainis \& Ivison
(2002) measure a 850~$\mu$ flux density of 8.1~mJy.

\smallskip

\noindent{\it CLASS B1933+503}

\smallskip

The lensed object in B1933+503 is a triple radio source (an MSO); two of
the components of the triple are quadruply imaged while the third is
doubly imaged (Sykes et al., 1998; Marlow et al., 2000). The source
shows rich milliarcsec-scale structure and, since many lines of sight
through the lensing galaxy are sampled, it is an excellent system on
which to test mass models (Nair, 1998: Cohn et al., 2001). An attempt
has been made to measure a time delay using VLA observations but was
unsuccessful owing to a lack of source variability (Biggs et al.,
2000). Two of the lensed images are detected with NICMOS (Marlow et
al., 1999b), the other two being presumed hidden due to extinction in
the lensing galaxy. Interestingly, the same two radio images appear to
suffer scatter broadening in the ISM of the lensing galaxy (Marlow et
al., 1999b). Chapman et al. (1999) report the detection of the system
at 850~$\mu$ and 450~$\mu$ with flux densities of 24~mJy and 114~mJy,
respectively, and suggest that the lensed object is a dusty quasar. A
redshift of 2.62 has been recently determined for the lensed object
(Norbury, Jackson \& Kerr, in preparation).

\smallskip

\noindent{\it CLASS B1938+666}

\smallskip

The discovery of B1938+666 was reported by King et al. (1997). It was
recognized as a lensed system by virtue of its extended arcs of radio
emission and therefore is not included in the statistically
well-defined sample of lens systems. The NICMOS image of the system
reveals an almost perfect Einstein ring with the lensing galaxy in the
centre (King et al., 1998). Barvainis \& Ivison (2002) measure
850~$\mu$ and 450~$\mu$ flux densities of 34.6~mJy and 126~mJy,
respectively.

\noindent{\it CLASS B2045+265}

\smallskip

Fassnacht et al. (1999a) reported the discovery of this 4-image lens
system. A fifth radio component is detected but it has a different
radio spectrum from the four others and is therefore likely to be
radio emission from the lensing galaxy itself. The lens has an optical
spectrum resembling that of an Sa galaxy (Fassnacht et al., 1999a). HST
observations show the lensed images and the lensing galaxy. Barvainis \& Ivison (2002) measure an upper limit of
3.7~mJy to the 850~$\mu$ flux density.

\smallskip

\noindent{\it CLASS B2108+213}

\smallskip

With an image separation of 4.6~arcsec, B2108+213 is the widest
separation lens system in the CLASS survey (McKean et al., in
preparation). Observations with the Keck Telescope show optical
counterparts to the radio images as well as the main lensing galaxy at
a redshift of 0.365 (Koopmans et al., in preparation).  A companion
galaxy within the Einstein radius is also detected and, possibly, a
cluster of faint red objects. As well as the two main radio images,
the strongest of which shows some faint extension in VLBA maps, there
is a third, 1.5~mJy radio component which is either a third image or
emission from the lensing galaxy. The latter interpretation is
favoured since the component seems to have an angular size of
$\sim$50~mas in a MERLIN 5~GHz map. The sum of the 8.4~GHz flux
densities of the images is $<$20~mJy, thus B2108+213 is not part of the
statistically well-defined sample.

\smallskip

\noindent{\it CLASS B2114+022} 

\smallskip

B2114+022 is the most enigmatic of the CLASS lens systems. Its
discovery is reported in Augusto et al. (2001) and lens modelling is
discussed in the companion paper by Chae, Mao \& Augusto (2001). The radio maps
show four compact components within 2.6~arcsec of each other but not
in an obvious lensing configuration. Two of the components have
inverted radio spectra and two have significantly steeper spectral
indices.  VLBI maps show that the former two components have a
much lower surface brightness than the latter supporting the view that
all four components cannot be images of the same object. Two potential
lensing galaxies are detected with redshifts of 0.3157 and 0.5883 but
no optical emission from any of the compact radio components has been
detected in either HST WFPC or HST NICMOS pictures. The most likely
scenario is that two images of a distant source are being detected
together with two radio components associated with the lower redshift
galaxy which has a starburst spectrum. Barvainis \& Ivison (2002)
measure an upper limit of 4.3~mJy to the 850~$\mu$ flux density.

\smallskip

\noindent{\it CLASS B2319+051} 

\smallskip

Rusin et al. (2001b) have described the lens system B2319+051. VLBA
observations resolve the two images into a pair of parity-reversed
sub-components separated by 20~mas in the stronger image and 7~mas in
the weaker. HST and ground-based optical observations reveal a primary
lensing galaxy at a redshift of 0.624 and another galaxy 3.4~arcsec
away with a redshift of 0.588 (Lubin et al., 2000). No optical
emission has been seen from the lensed images. Barvainis \& Ivison
(2002) measure 850~$\mu$ and 450~$\mu$ flux densities of 3.9~mJy and
40~mJy, respectively.

\section{Discussion}

In this section we discuss some of the basic statistical properties of
the CLASS lens systems. First, however, we reiterate the numbers of
objects in the different CLASS subsamples since these numbers are
fundamental to the calculation of the point-source lensing rate. The
CLASS complete sample of flat-spectrum radio sources, derived from GB6
and NVSS contains 11,685 objects. However, to be on the list from
which the 149 lens candidates were selected these sources need to
satisfy additional conditions (See Table 1), most importantly having a
total detected 8.4~GHz flux density $\>$20~mJy. {\it Hence the CLASS
statistical sample used in the subsequent analysis and that of Paper 3
consists of 8,958 objects.}

\subsection{The overall lensing rate and the relative numbers of quads and doubles}

Our search for multiply-imaged gravitational lens systems in a
well-defined sample of flat-spectrum radio sources is complete in the
specified range of parameter space; i.e. image separation between 0.3
and 15~arcsec and flux density ratio $\leq$10:1.  There are 13 lens
systems meeting these criteria amongst the 8,958 sources in the
well-defined parent sample. Thus the point-source lensing rate is 1 in
690$\pm$190 with the error being derived on the basis of Poisson
statistics. The lensing rate amongst the remaining, less tightly
selected, CLASS targets is similar (9 out of $\sim$5200;
i.e. 1:600). This lensing rate depends on the selection criteria; we
miss some systems because we restrict the image separations to be
$\geq$0.3~arcsec and some because we require the image flux density
ratios for double systems to be $\leq$10:1.  We have made predictions
of the effect of these criteria assuming the best-fit models in flat
Universe models found in Paper 3.  Ignoring image separations
$<$300~mas will lead us to miss between 13 and 17 \% of the systems,
the exact value depending somewhat on the details of the luminosity
functions adopted for the spiral and elliptical lensing galaxies.
Restricting our search to separations $>$300~mas systematically
selects against spiral lenses (Turner, Ostriker \& Gott, 1984; Augusto
\& Wilkinson, 2001).

The percentage of double systems
\footnote{The strongest images of quad systems always have flux
density ratios $\sim<$3:1; Shude Mao, personal communication.}  missed 
due to ignoring candidates with
image flux density ratios $>$10:1 is expected to be $\sim$37\%, the
exact value depending on the distribution of ellipticities in the
lensing galaxy population. This calculation assumes a slope of -2.07
for the differential number/flux density relation above 30~mJy and a
slope of -1.97 below 30~mJy.  We know of one CLASS system B1030+074
with a flux density ratio $>$10:1 which has been excluded from the
statistically well-defined lens sample for this reason. Since we have
followed up many other high flux-ratio candidates without finding any
more such lens systems, we regard 37\% as an upper limit to the
percentage of double systems excluded for this reason.

It is also interesting to look at the relative numbers of 5-image
systems (seen as quads) and of 3-image systems (seen as doubles). In
the statistically well-defined sample there are six
quads\footnote{Actually there are five quads plus one 6-image system
B1359+154 (Myers et al 1999; Rusin et al 2001a)} and seven doubles. If
we consider all the 22 systems found in CLASS then we find 10 quads
and 12 doubles.  This is a higher fraction of quads than the
$\sim$0.25 expected on the basis of simple models (King \& Browne,
1996; Kochanek, 1996) and is also higher than found amongst non-CLASS
lens systems. For example, amongst the non-CLASS lenses listed on the
CASTLES web page (http://cfa-www.harvard.edu/castles/; Falco et al.,
2001) there are nine quads and 32 doubles. However, the CLASS results
are likely to be more reliable because they do not have the unknown
selection biases that are suffered by the heterogeneous group of known
lens systems. Rusin and Tegmark (2001) have looked at the CLASS
statistics and consider a range of factors (e.g. external shear
fields, mass distributions flatter than the light, shallow lensing
mass profiles, finite core radii, satellite galaxies, etc.)  which may
increase the frequency of radio quads. They conclude that none of the
mechanisms provide a compelling solution to the problem. In Paper 3 it
is concluded that the average ellipticity of the lensing masses has to
be $\geq$0.17 at 95\% confidence level in order to fit the observed
CLASS lensing statistics.

\subsection{Missing doubles and multi-path scattering}

Could we be systematically missing double lens systems? We believe
not. The only way we can suggest that this might conceivably happen is
if the importance of a propagation effect such as multi-path
scattering occurring in the ionised ISM of lensing galaxies has been
underestimated. Scattering would have the effect of increasing the
angular size of an image, and hence decreasing its measured surface
brightness. Thus there would be a chance of rejecting a genuine system
for the reason that the measured image surface brightnesses were
different. This might be a problem only for the 5~GHz VLBA
observations which have a resolution of $\sim$3~mas. Multi-path
scattering occurring in the lensing galaxy has been invoked as a
possibility in B1933+503 (Marlow et al., 1999b), in PKS 1830-211
(Guirado et al., 1999), in B2114+024 (Augusto et al., 2001) and in
B0218+357 (Biggs et al., 2002). However, B0128+437 (Biggs et al., in
preparation) is the only possible case we know amongst the $\sim$60
lines of sight probed by CLASS core images in which an image surface
brightness appears to have been changed enough to potentially affect
our conclusions based on the 5~GHz VLBA observations. It is a quad
system in which three of the images have a clearly visible compact
core component, whereas the fourth does not. Since it is a quad, there
was no difficulty in recognizing it as a lensed system but, if one of
the images of a double system had sampled the same line of sight
through the lensing galaxy, it might well have been rejected. The size
of a scattered image is expected to have a roughly $\lambda^{2}$
dependence and therefore high frequency observations should be much
less affected.  We plan 15~GHz VLBA observations of $\sim$10
candidates that have been rejected on the basis of surface brightness
differences in VLBA 5~GHz maps in order to eliminate any residual
doubt about the completeness of the CLASS search for double-image lens
systems.

\subsection{Dark lenses?}

Using preliminary CLASS results on 13 lensed systems Jackson et al.
(1998b) argued that the detection of the lensing galaxy with the HST
rules out the existence of a significant number of dark lenses capable
of producing image splittings of $\geq$1~arcsec, as suggested by
Hawkins (1998). We are now in a position to update this discussion
given that there are now optical/infrared observations of sufficient
quality to separate lensed images and lensing galaxies of 20 of the 22
CLASS lens systems. (For B0445+123 only ground-based imaging is
currently available and for B1555+375, existing HST imaging has not
yet unambiguously revealed a lensing galaxy.)  In all 20 cases the
lensing galaxy is detected. We stress that we have not used the
absence of a lensing galaxy to rule out lensing. There is just one
case in where a candidate has passed all the basic radio tests and no
lensing galaxy has been detected. J0831+524 (B0827+525 -- Koopmans et
al., 2000a) was discussed in Section 4.1.2 and is the only possible dark lens
candidate in CLASS. The accumulated evidence is against the lensing hypothesis
although further observations are planned. Thus the conclusion of
Jackson et al. (1998b) that galaxy-mass dark lenses producing image
separations in the range 0.3 to 15 arcsec are rare is reinforced by
the complete CLASS results.

\subsection{Image separations}

\begin{figure*}
\centering \setlength{\unitlength}{1in}
\begin{picture}(7.5,6.5)
\put(1.6,0.3){\includegraphics{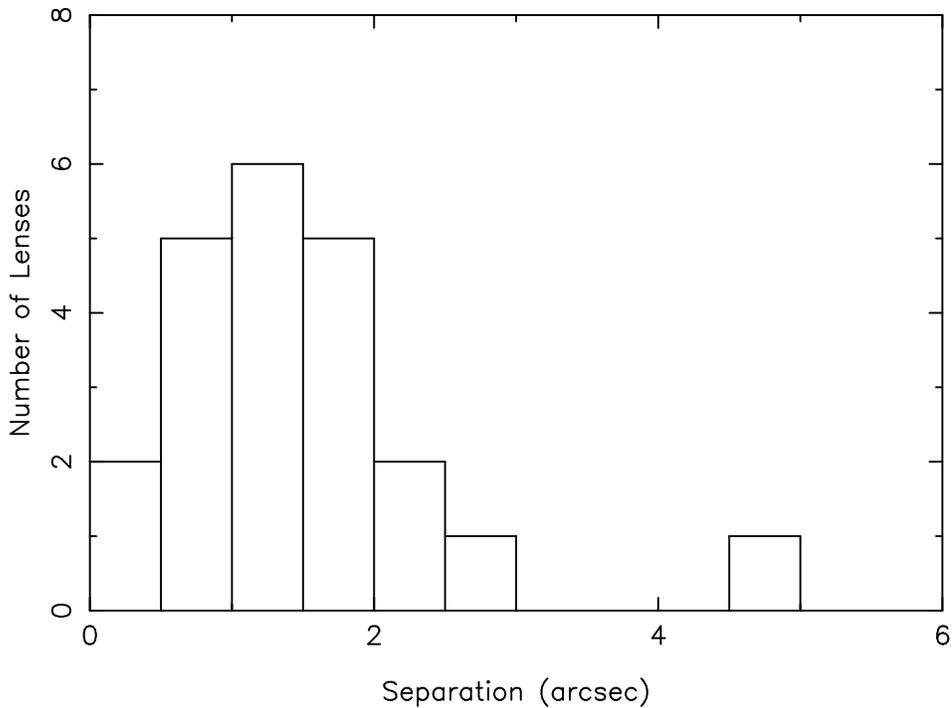}}
\end{picture}
\caption{Separation histogram for the 22 confirmed CLASS lens systems. It
should be noted that the contents of the smallest separation bin between 0 and
0.5~arcsec may be affected by the survey restriction to images separations $\leq$0.3~arcsec.
\label{separations}}
\end{figure*}

A histogram of image separations for all CLASS lens systems is shown
in Figure~\ref{separations}.  The separations peak at just over
1~arcsec, not very different from the predictions of Turner, Ostriker
\& Gott (1984). However, it should be noted that there is evidence for
multiple lensing galaxies in several of the lens systems (B1127+385,
B1359+154, B1608+656, B2108+213 and B2114+022) suggesting that the
characteristic image separation for lensing by a single galaxy may
be $\leq$1~arcsec.

Only one lens system, B2108+213, has a maximum image separation
$\geq$3~arcsec and that is ``cluster assisted''; i.e. the lens
separation is affected by the presence of several other galaxies presumed to
belong to the same cluster or group as the dominant lensing galaxy. In
this it resembles the first gravitational lens system to be discovered
B0957+561 (Walsh, Carswell \& Weymann, 1979).  We have searched for
systems with image separations from 6 up to 15~arcsec and found none
(Phillips et al., 2002). Unsuccessful searches for systems with
separations $\leq$0.3~arcsec have been made (Augusto \& Wilkinson,
2001; Wilkinson et al., 2001) but with significantly fewer objects
than in the CLASS sample. Nevertheless they are sufficient to show
that the lensing rate for systems $\le$0.3~arcsec is lower than that
for $\sim$1~arcsec with a high degree of confidence. The peak in the
histogram is real.

\bigskip

\subsection{Time delay and the Hubble constant}

One of the aims of JVAS and CLASS was to identify lens systems
suitable for time delay measurements and Hubble constant
determination. The seven systems which have been monitored
systematically for time delays are listed in Table~\ref{delays}. Four
systems have measured time delays, three of them (B0218+357, B1600+434
and B1608+656) with sufficient accuracy to lead to useful Hubble
constant determinations in the region of $\sim$70~km s$^{-1}$ Mpc$^{-1}$. Koopmans \& Fassnacht (1999) summarized the
various gravitational lens H$_{0}$ determinations at that
time. Refined values based on analyses of new observations of
B0218+357, B1600+434 and B1608+656 are likely to be available shortly.


\begin{table*}
\caption{Gravitational lenses in the CLASS sample which have been monitored for time delays.
\label{delays}}
\begin{tabular}{lll}
\hline 
source  & references  & time delay     \\
\hline
B0218+357  & Biggs et al., 1999 & B and A  10.4$\pm$0.4~days   \\
B0414+054  & Moore \& Hewitt, 1997 & --     \\
B1030+074  & Xanthopoulos, unpublished & -- \\
B1422+231  & Patnaik \& Narasimha, 2001 & B and A 1.5$\pm$1.4~days \\
           &                            & A and C 7.6$\pm$2.5~days \\
           &                            & B and C 8.2$\pm$2.0~days  \\ 
B1600+434  & Koopmans et al., 2000b  & B and A 47$\pm$6~days    \\
B1608+656  & Fassnacht et al., 2002 & B and A 31.5$\pm$3~days   \\
           &                        & B and C 37$\pm$3~days   \\
           &                        & B and D 76$\pm$3~days    \\
B1933+503  & Biggs et al., 2000 & --   \\
\hline
\end{tabular}
\end{table*}

\section{conclusions}

The CLASS methodology, using high resolution radio observations of
flat-spectrum radio sources to define reliable and complete samples of
gravitational lens systems, works efficiently. Many of the lens
systems turn out to be suitable for time delay measurements; time
delays have been measured for three lens systems found in the survey
and more will follow. The point-source lensing rate is 1:690 $\pm$190
targets. The discussion of the implications of the lensing statistics
for cosmology and galaxy evolution is discussed in Paper 3.

\section*{Acknowledgements}
We thank Shude Mao for many insightful discussions. The National Radio Astronomy Observatory is a facility of the National
Science Foundation operated under cooperative agreement by Associated
Universities, Incorporated. MERLIN is operated by the University of
Manchester as a National Facility of the Particle Physics \& Astronomy
Research Council. This work was supported by the European Commission,
TMR Programme, Research Network Contract ERBFMRXCT96-0034
``CERES''. This research has made use of the NASA/IPAC Extragalactic
Database (NED) which is operated by the Jet Propulsion Laboratory,
California Institute of Technology, under contract with the National
Aeronautics and Space Administration.


\end{document}